\documentclass[iop]{emulateapj}
\begin{document}

\slugcomment{Accepted to the Astrophysical Journal}

\shorttitle{Atmospheric circulation of GJ 1214b}
\shortauthors{Kataria et al.}

\title{The atmospheric circulation of the super Earth GJ 1214b: Dependence on composition and metallicity}

\author{T. Kataria, A.P. Showman}
\affil{\it{Department of Planetary Sciences and Lunar and Planetary Laboratory, The University of Arizona,
Tucson, AZ 85721}}
\email{\bf{tkataria@lpl.arizona.edu}}

\author{J.J. Fortney}
\affil{\it{Department of Astronomy \& Astrophysics, University of California, Santa Cruz, CA 95064}}

\author {M.S. Marley}
\affil{\it{NASA Ames Research Center 245-3, Moffett Field, CA 94035}}

\and
\author {R.S. Freedman}
\affil{\it{SETI Institute, 189 Bernardo Ave \#100, Mountain View, CA 94043}}
\affil{\it{NASA Ames Research Center 245-3, Moffett Field, CA 94035}}

\begin{abstract}
We present three-dimensional atmospheric circulation models of GJ 1214b, a 2.7 Earth-radius, 6.5 Earth-mass super Earth detected by the MEarth survey.  Here we explore the planet's circulation as a function of atmospheric metallicity and atmospheric composition, modeling atmospheres with a low mean-molecular weight (i.e., $\mathrm{H_2}$-dominated) and a high mean-molecular weight (i.e. water- and $\mathrm{CO_2}$-dominated).  We find that atmospheres with a low mean-molecular weight have strong day-night temperature variations at pressures above the infrared photosphere that lead to equatorial superrotation.  For these atmospheres, the enhancement of atmospheric opacities with increasing metallicity lead to shallower atmospheric heating, larger day-night temperature variations and hence stronger superrotation.  In comparison, atmospheres with a high mean-molecular weight have larger day-night and equator-to-pole temperature variations than low mean-molecular weight atmospheres, but differences in opacity structure and energy budget lead to differences in jet structure.  The circulation of a water-dominated atmosphere is dominated by equatorial superrotation, while the circulation of a $\mathrm{CO_2}$-dominated atmosphere is instead dominated by high-latitude jets.  By comparing emergent flux spectra and lightcurves for 50$\times$ solar and water-dominated compositions, we show that observations in emission can break the degeneracy in determining the atmospheric composition of GJ 1214b.  The variation in opacity with wavelength for the water-dominated atmosphere leads to large phase variations within water bands and small phase variations outside of water bands.  The 50$\times$ solar atmosphere, however, yields small variations within water bands and large phase variations at other characteristic wavelengths.  These observations would be much less sensitive to clouds, condensates, and hazes than transit observations.
\end{abstract}


\section{Introduction}
As the number of extrasolar planets detected by various ground- and space-based surveys grows, so too do the number of so-called ``super Earths", exoplanets with masses of 1-10 Earth masses.  Many of these super Earths transit their host stars along our line of sight, which allow us to directly observe their atmospheres using the same techniques as for hot Jupiters \citep[e.g.,][]{redfield2008}.  Such a case is true for GJ 1214b, a  2.7 Earth-radius, 6.5 Earth-mass super Earth detected by the MEarth survey \citep{charb2009}.  Because GJ 1214A is an M-type star only 13 parsecs away, the system has proven to be a favorable target for follow-up observations \citep[e.g.,][]{bean2010, bean2011, berta2012, croll2011, crossfield2011, narita2012, demooij2012, fraine2013, teske2013, kreidberg2014}.  

\cite{charb2009} concluded that the measured mass and radius of GJ 1214b is most consistent with an interior that is water-dominated, with a hydrogen-helium envelope that is 0.05\% the mass of the planet.  \cite{rs2010} modeled the planet's interior structure, and concluded that if water were present in the planet's atmosphere, it would be a supercritical fluid.  Hence, GJ 1214b should not have a solid surface.  \cite{nettelmann2011} also modeled the interior of GJ 1214b assuming a two-layer (homogeneous envelope overlying a rock core) structure, and found their results favor a composition similar to that of \cite{charb2009}.  \cite{va2013} ran a range of internal structure/evolution models (H/He or $\mathrm{H_{2}O}$ envelope overlying an Earth-like nucleus), finding that only a small amount of H/He is needed to explain the planet's mass and radius.

In anticipation of follow-up observations of GJ 1214b, \cite{mrf2010} modeled transmission and emission spectra for a range of atmospheric compositions, from hydrogen-dominated (i.e., those with a low mean-molecular-weight [MMW] atmosphere) to $\mathrm{CO_2}$- and $\mathrm{H_2 O}$-dominated (i.e., those with a high mean-molecular weight).  They found that if the planet's atmosphere were $\mathrm{H_{2}/He}$-dominated, the primary transit depth would show larger variations with wavelength than if the planet had an $\mathrm{H_2 O}$- or $\mathrm{CO_2}$-dominated atmosphere; this is because of the larger atmospheric scale height for an $\mathrm{H_2}$-dominated atmosphere as compared to a high-MMW atmosphere.  This would lead to enhanced spectral features that should be detectable by current  ground- and space-based instrumentation.  

Transit spectroscopic observations by most groups, however, favor a flat transmission spectrum, consistent with a high-MMW (e.g., water) atmosphere or an atmosphere with high-altitude clouds or hazes \citep[e.g.,][]{bean2010, bean2011, berta2012, narita2012, demooij2012, fraine2013}.  Still, observations by other groups favor a low-MMW atmosphere \citep{croll2011}, particularly if methane is depleted \citep{crossfield2011}.  Photochemical modeling by \cite{mrk2012} also support a methane depletion, consistent with methane photolysis, but note that this process is not efficient at the pressure levels probed by transmission spectroscopy.  

The composition will affect not only the atmospheric opacities (hence absorption of starlight and emission of infrared radiation) but also the atmospheric scale height, dry adiabatic lapse rate, and hence the dynamical stability and circulation of the atmosphere.  The circulation will determine the location of hot and cold regions in the atmosphere, which, in turn, shapes lightcurve and spectral behavior at photospheric levels. In light of these considerations, we model the atmospheric circulation of GJ 1214b, testing a multitude of atmospheric compositions.  The circulation of GJ 1214b has been explored by other groups \citep{menou2012, zalucha2012}.  However, our circulation model incorporates the most rigorous radiative transfer scheme used to model the atmosphere thus far (see below).  In Section 2, we describe our general circulation model, the SPARC/MITgcm, and describe our model integrations.  In Section 3, we present results from our model integrations, and identify general trends in circulation and temperature structure with metallicity and composition.  In Section 4, we generate emergent flux spectra and lightcurves in anticipation of future instrumentation aboard the {\it James Webb Space Telescope} (JWST), {\it Thirty Meter Telescope} (TMT) and other other ground- and space-based facilities.

\section{Model}

\begin{figure}
\begin{centering}
\epsscale{.80}
\includegraphics[trim = 0.7in 2.8in 1.0in 2.5in, clip, width=0.40\textwidth]{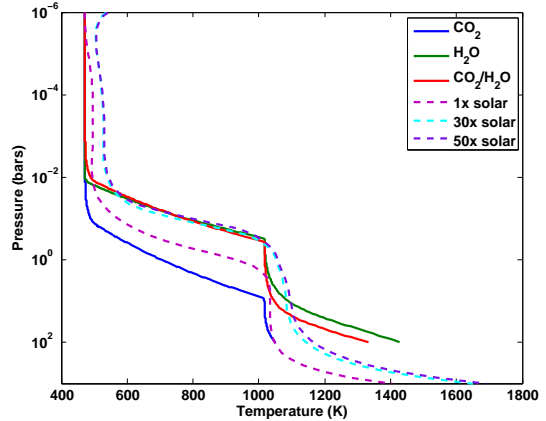}\\
\caption{1-D initial pressure-temperature (P-T) profiles for our model integrations, adapted from \cite{mrf2010}. Each profile assumes $4\pi$ redistribution of incident sunlight.}
\label{pt_plots}
\end{centering}
\end{figure}

\subsection{The SPARC/MITgcm}
The atmospheric circulation of GJ 1214b is modeled using the Substellar and Planetary Atmospheric Radiation and Circulation (SPARC) Model \citep{showman2009}, which couples the MITgcm, a general circulation model (GCM) maintained at the Massachusetts Institute of Technology \citep{adcroft2004}, with a two-stream implementation of the multi-stream, non-gray radiative transfer scheme developed by \cite{mm1999}.  To emphasize its heritage, we refer to this model as the SPARC/MITgcm.  The MITgcm solves the primitive equations, a simplification of the Navier-Stokes equations assuming local hydrostatic balance, which is valid in stably stratified atmospheres with horizontal length scales greatly exceeding vertical length scales.  The primitive equations are solved on a cubed sphere grid, allowing for longer timestepping and better accuracy near the poles as compared to a latitude-longitude grid.  The radiative transfer code solves for the upward and downward fluxes through a given vertical column of atmosphere in the GCM, which determines the heating rate used to update the temperature and winds.  For each chosen atmospheric composition (see below) the opacities are divided into 11 frequency bins using the correlated-$k$ method (Goody et al. 1989; for more details on the SPARC/MITgcm, including recent updates to the model, see Showman et al. 2009 and Kataria et al. 2013).  Each model integration has a horizontal resolution of C32 ($\sim64\times$128 in latitude and longitude) and 40 or 76 pressure levels.  The  pressure levels extend from a mean pressure of 200 bars at the bottom to 0.2 mbar at the top, evenly spaced in log pressure.  The top level extends from a pressure of 0.2 mbar to zero.

The SPARC/MITgcm has been successfully adapted to investigate a variety of aspects of the atmospheric dynamics of hot Jupiters and hot Neptunes \citep{showman2009, showman2013, lewis2010, parmentier2013, kataria2013}.  While the MITgcm is classically an Earth GCM, this is the first time the SPARC/MITgcm in its entirety has been used to model the circulation of a super Earth.  However, given the likelihood that GJ 1214b does not have a solid surface based on its mass, radius and temperature, we can use the SPARC/MITgcm with few adjustments.  Utilizing the SPARC/MITgcm for rocky, terrestrial exoplanets will be a task for future studies.

\begin{deluxetable*}{lccc}
\tabletypesize{\scriptsize}
\tablecaption{Molecular weight, Specific heat ($c_p$) and Scale height ($H$) values for each atmospheric composition. \label{cp_values}}
\tablewidth{0pt}
\tablehead{
\colhead{Atmospheric composition} & \colhead{$c_p$ ($\mathrm{J~kg^{-1}~K^{-1}}$)} & \colhead{$H$ ($\mathrm{km}$)} & \colhead{Mean-molecular weight ($\mathrm{g~mol^{-1}}$)}
}
\startdata
1$\times$ solar & 13000 & 230 & 2.228\\
30$\times$ solar & 9440 & 175 & 2.936\\
50$\times$ solar & 8213 & 150 & 3.424\\
$\mathrm{H_2 O}$-dominated (99\% $\mathrm{H_2O}$, 1\% $\mathrm{CO_2}$) & 1981 & 28 & 18.026\\
$\mathrm{CO_2}$-dominated (99\% $\mathrm{CO_2}$, 1\% $\mathrm{H_2O}$) & 1016 & 12 & 43.974\\
$50\%~\mathrm{CO_2}$, $50 \%~\mathrm{H_2 O}$ & 1296 & 17 & 31.00\\
\enddata
\label{rundata}
\end{deluxetable*}

\subsection{Model integrations}
We model six atmospheric compositions for GJ 1214b, adapted from \cite{mrf2010}.  First, we model $\mathrm{H_2}$-dominated (i.e., low-MMW) compositions at $1\times, 30\times$, and $50\times$ solar, which have mean molecular weights of 2.228, 2.936, and 3.424 $\mathrm{g~mol^{-1}}$, respectively.  These models assume molecular species are in chemical equilibrium abundances at the local temperature and pressure, accounting for rainout of species that have condensed.  For the high metallicity cases, all species except for $\mathrm{H_2/He}$ are enhanced by their respective factors.  Second, we model an $\mathrm{H_{2}O}$-dominated atmospheric composition, which is composed of 99\% $\mathrm{H_2O}$, and 1\% $\mathrm{CO_2}$.  Third, we model a $\mathrm{CO_2}$-dominated atmospheric composition (99\% $\mathrm{CO_2}$, 1\% $\mathrm{H_2O}$).  Lastly, we model an intermediate high-MMW case, with a composition of $\mathrm{50\%~CO_2}$ and $\mathrm{50 \%~H_2 O}$.  

For each model integration, we assume the winds to be initially zero, and assign each vertical atmospheric column the global-mean radiative-equilibrium temperature-pressure profile calculated using a one-dimensional (1-D) radiative transfer code.  \cite{liu2013} have shown that hot, synchronously rotating exoplanets exhibit circulation patterns that are insensitive to initial conditions.  Figure \ref{pt_plots} shows the pressure-temperature (P-T) profiles used in these initial conditions.  The hydrogen-dominated 1-D P-T profiles were calculated using the radiative-transfer code of \cite{fortney2005,fortney2006,fortney2008} adapted from \cite{mm1999}.  The $\mathrm{H_2 O}$- and $\mathrm{CO_2}$-dominated 1-D profiles were generated using the code of \cite{mr2009}.  Both codes calculate the temperature structure self-consistently assuming radiative equilibrium.  The SPARC/MITgcm self-consistently solves for the flow as dynamics and heating evolve.  

In changing the atmospheric composition, we are also changing the mean-molecular weight, the specific heat, and the scale height.  We calculate the specific heat using the method described in \cite{cs2006}.  This is given on a per mass basis as

\begin{equation}\label{specificheat}
    c_p=c_{p1} \cdot X_1 + c_{p2} \cdot X_2 + \cdots + c_{pn} \cdot X_n 
\end{equation}

\noindent where $c_{pn}$ and $X_n$ are the specific heat and mixing ratio of the $n^{th}$ atmospheric constituent, respectively.  

The scale height, $H$, is given by $H=R_{s}T/g$, where $R_s$ is the specific gas constant, $T$ is the effective temperature, and $g$ is the planetary gravity.  The values of molecular mass, $c_p$ and $H$ for each composition are listed in Table \ref{rundata}, and vary over an order of magnitude.  

For each simulation, we use a dynamical timestep of 25 or 10 seconds with a radiative timestep of 500 or 200 seconds.  The simulations were each run for approximately 5000 Earth days, with outputs every 100 days.  

\section{Results}

\subsection{Hydrogen-dominated atmospheric composition}

\begin{figure*}
\begin{centering}
\epsscale{.80}
\includegraphics[trim = 0.0in 0.0in 0.0in 0.0in, clip, width=0.45\textwidth]{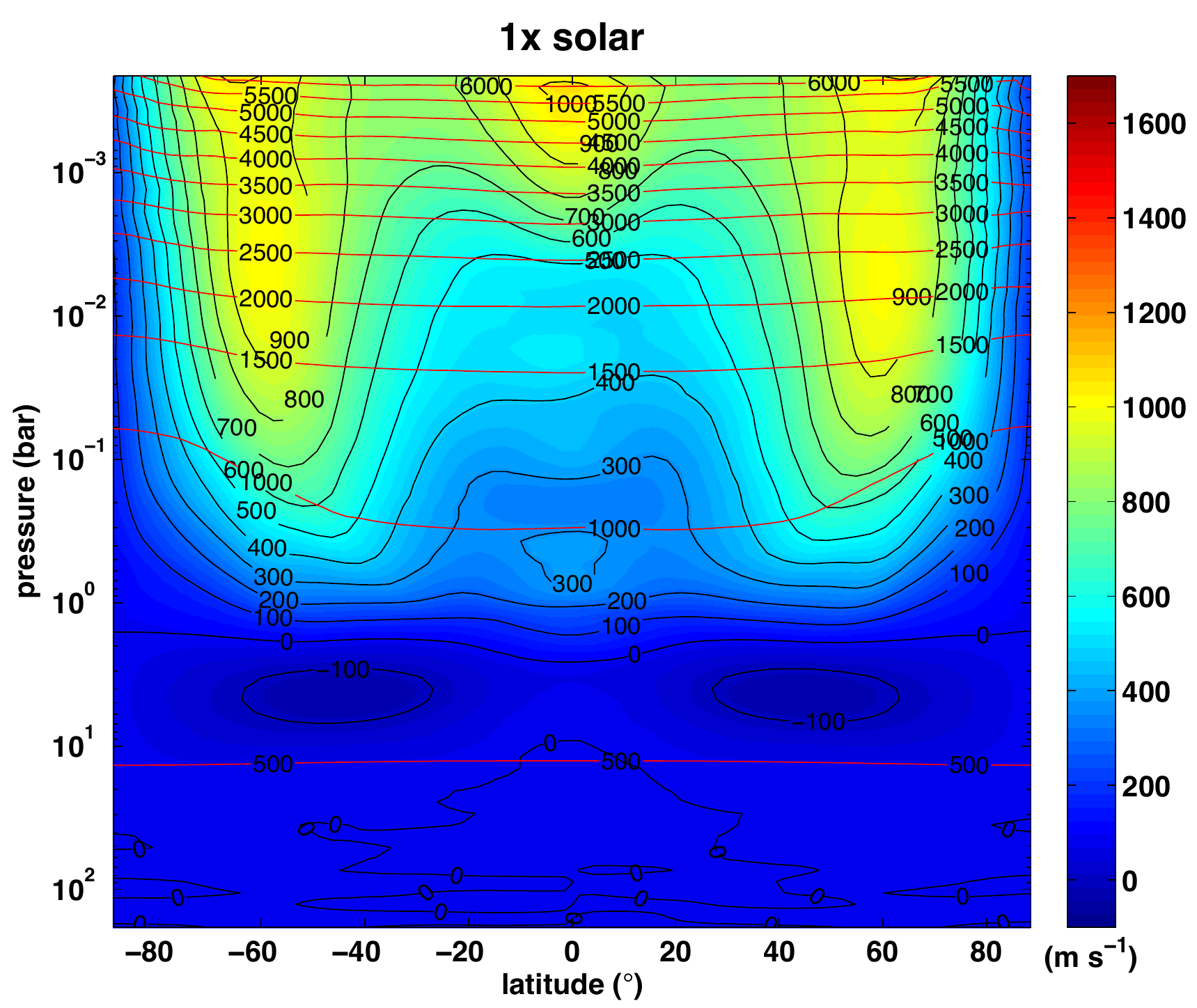}
\includegraphics[trim = 0.0in 0.0in 0.0in 0.0in, clip, width=0.45\textwidth]{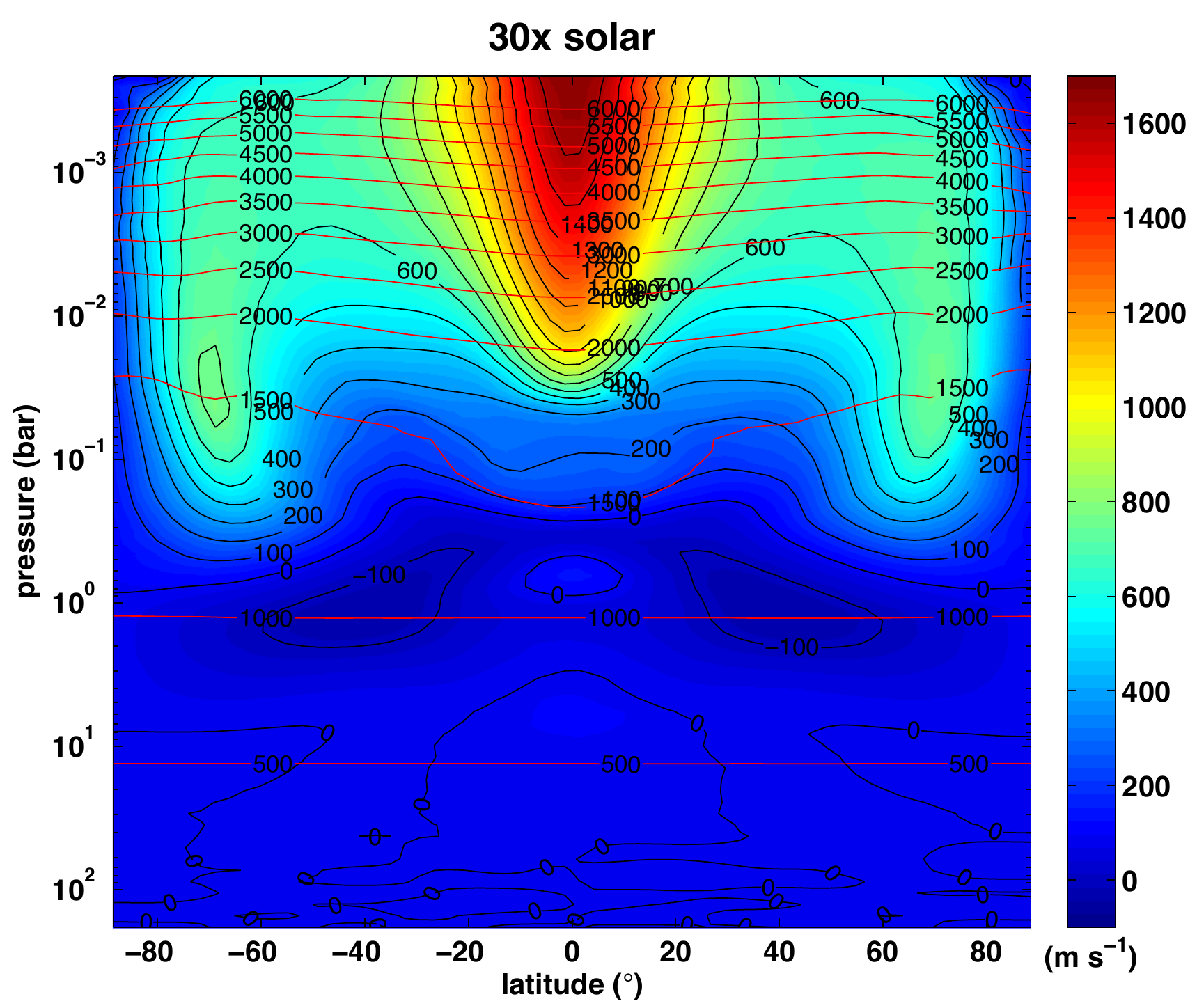}
\includegraphics[trim = 0.0in 0.0in 0.0in 0.0in, clip, width=0.45\textwidth]{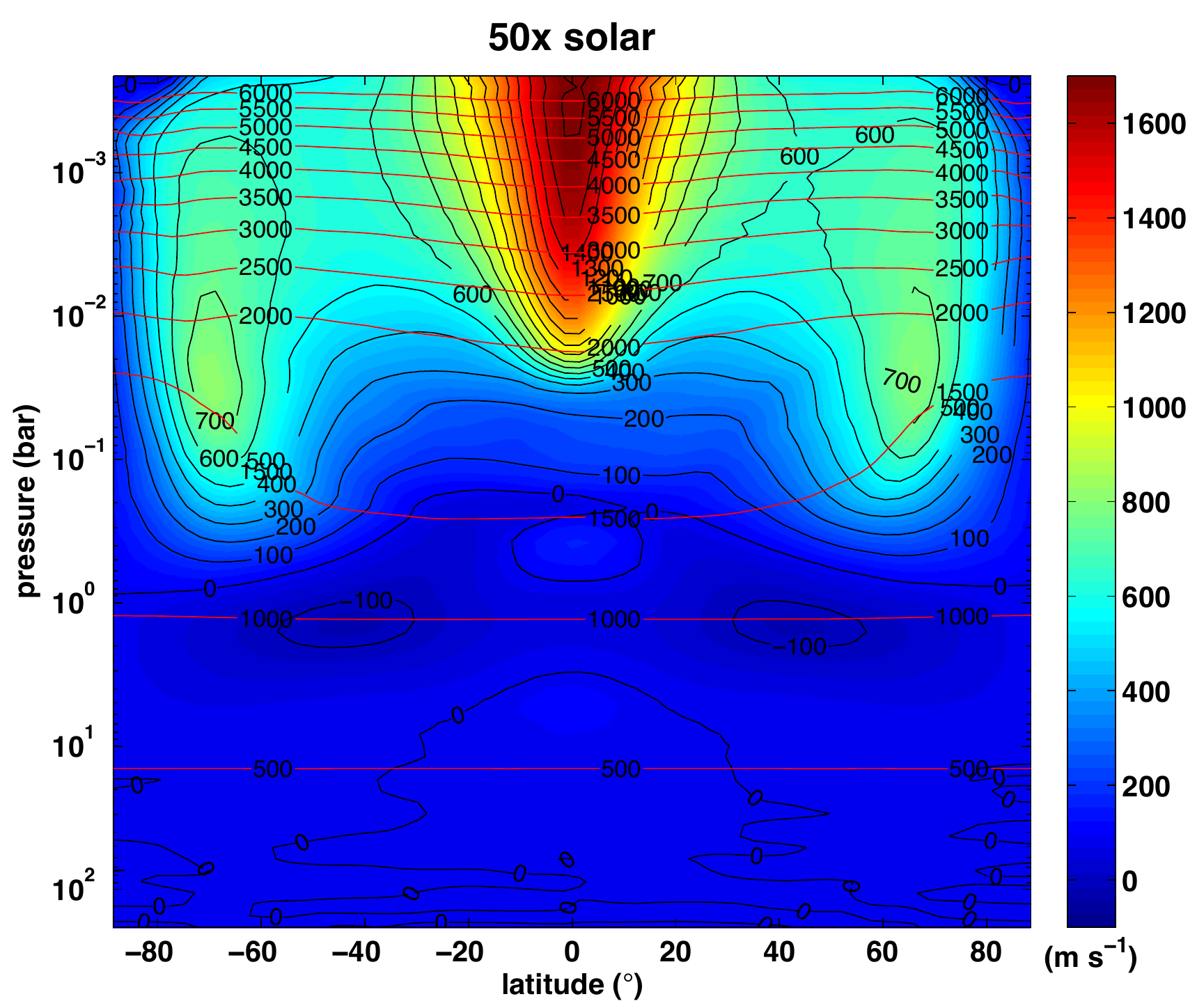}
\caption{Zonal-mean zonal wind for $\mathrm{H_2}$-dominated compositions of GJ 1214b.  The plots correspond to atmospheric compositions of 1$\times$, 30$\times$, and 50$\times$ solar. Zonal-mean isentropes (potential temperature contours) are overplotted in red in intervals of 500 K.  Note the winds are plotted on the same colorscale.}
\label{lowmmw_plots}
\end{centering}
\end{figure*}

For all three $\mathrm{H_2/He}$-dominated models, the atmospheres possess an equatorial superrotating jet, with speeds exceeding $1~\mathrm{km~s^{-1}}$.  Each model also exhibits a pair of jets in the high-latitudes.  This is seen in Figure \ref{lowmmw_plots}, which plots the zonal-mean zonal wind\footnote{The zonal wind is defined as the east-west wind, where positive (negative) values denote an eastward (westward) wind; a zonal mean denotes an average in longitude.  All zonal means are averaged in longitude along surfaces of constant pressure.  } averaged over a planetary orbit for the 1$\times$, 30$\times$ and 50$\times$ solar composition.  Overplotted in red are zonal-mean isentropes, contours of constant potential temperature.  For the 1$\times$ solar case, the high-latitude jets are centered at roughly $60^{\circ}$, with peak speeds comparable those at the equator.  For the high metallicity cases, the high-latitude jets are centered at $\sim70^{\circ}$, with speeds of 700 $\mathrm{m~s^{-1}}$.

\begin{figure*}
\begin{centering}
\epsscale{.80}
\includegraphics[trim = 0.0in 0.0in 0.0in 0.0in, clip, width=0.32\textwidth]{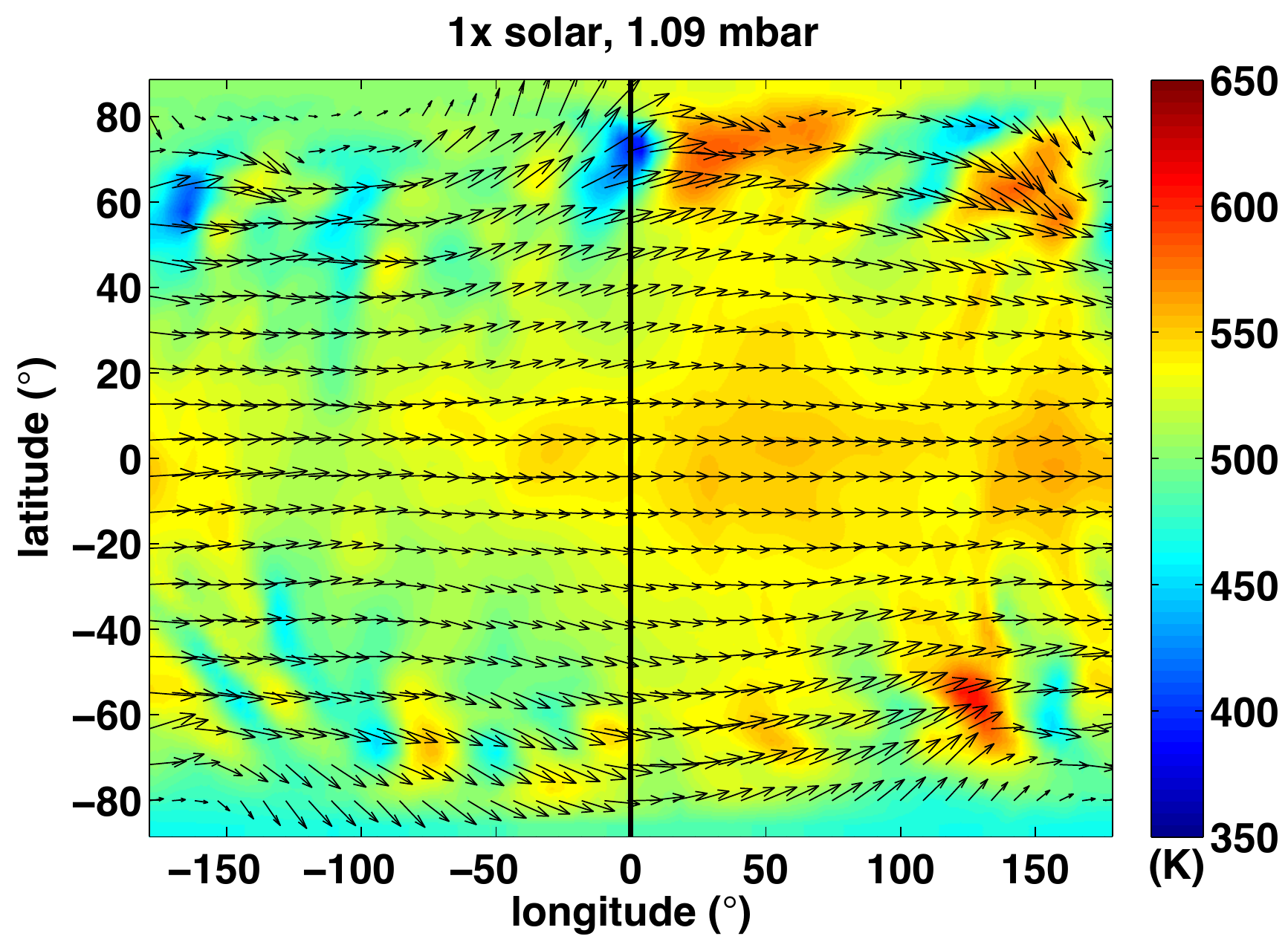} 
\includegraphics[trim = 0.0in 0.0in 0.0in 0.0in, clip, width=0.32\textwidth]{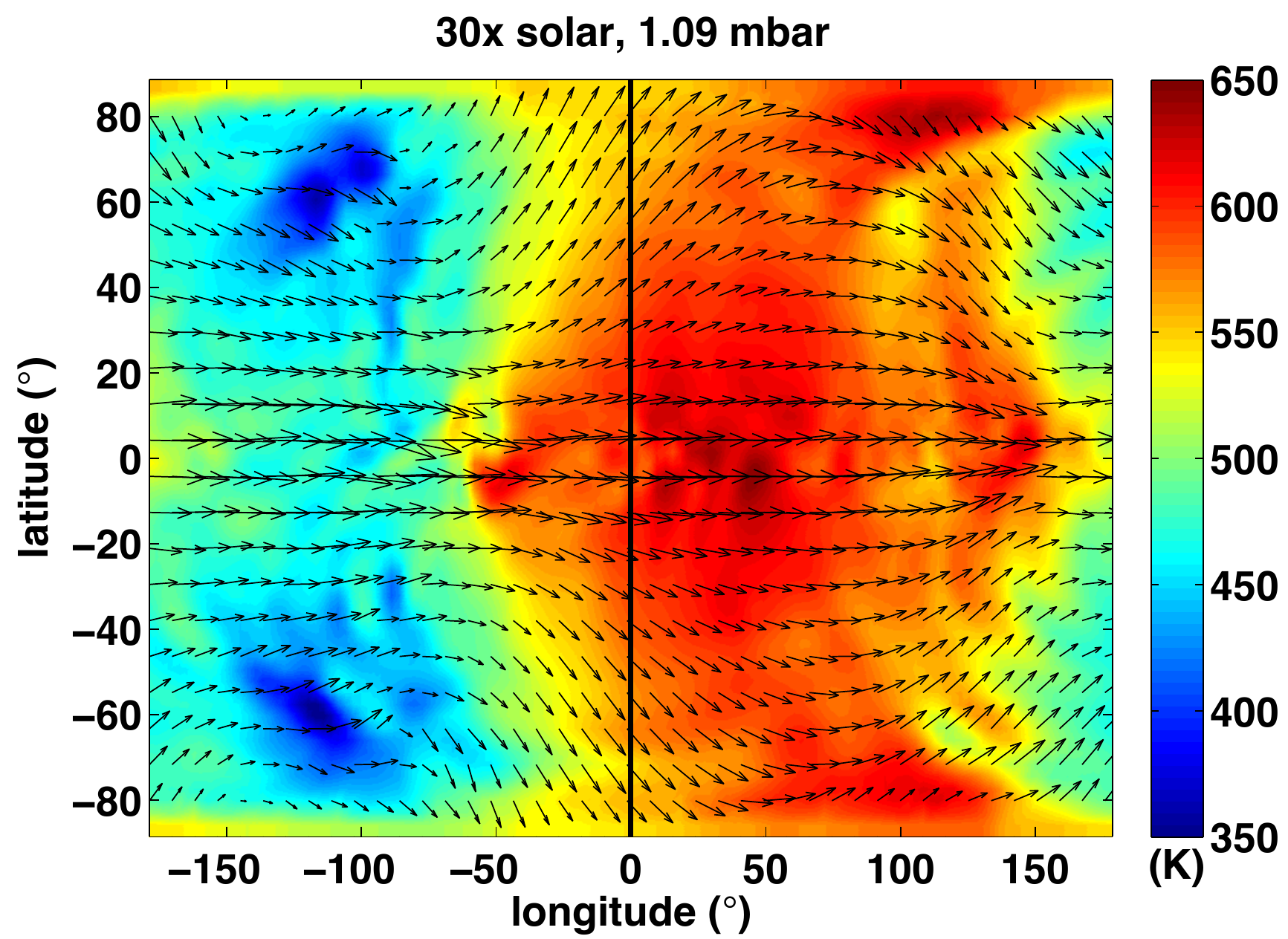} 
\includegraphics[trim = 0.0in 0.0in 0.0in 0.0in, clip, width=0.32\textwidth]{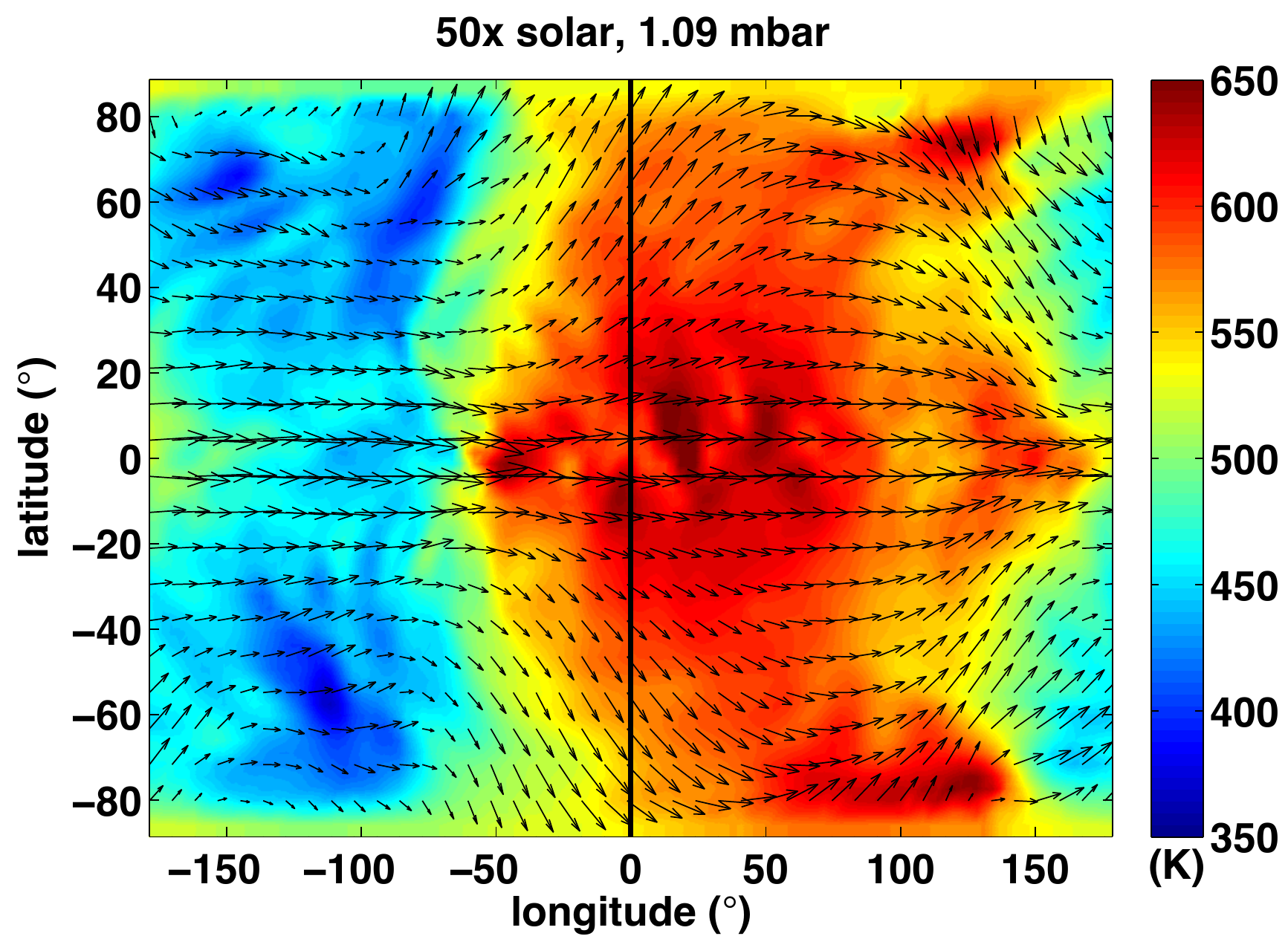}\\ 
\includegraphics[trim = 0.0in 0.0in 0.0in 0.0in, clip, width=0.32\textwidth]{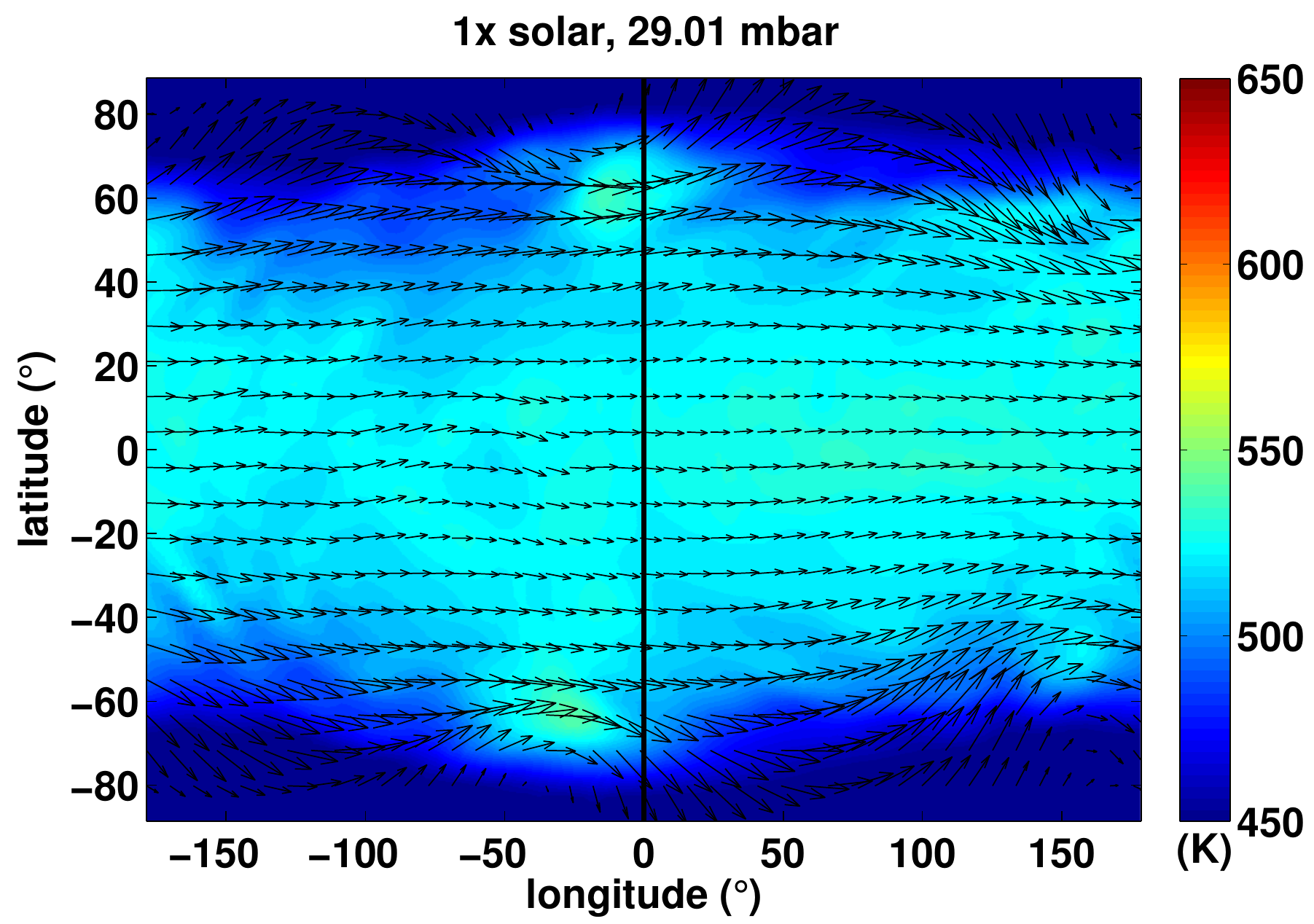} 
\includegraphics[trim = 0.0in 0.0in 0.0in 0.0in, clip, width=0.32\textwidth]{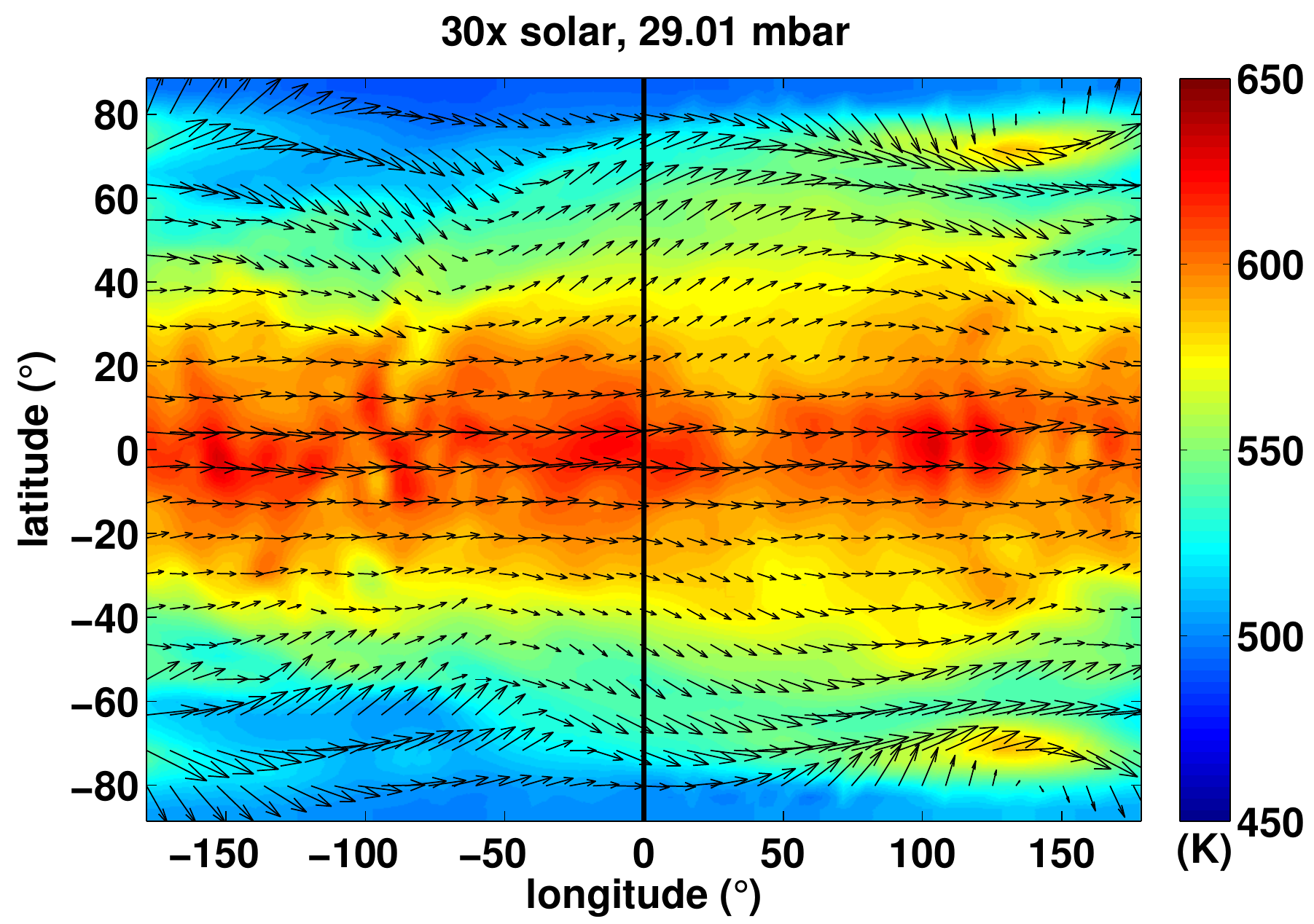} 
\includegraphics[trim = 0.0in 0.0in 0.0in 0.0in, clip, width=0.32\textwidth]{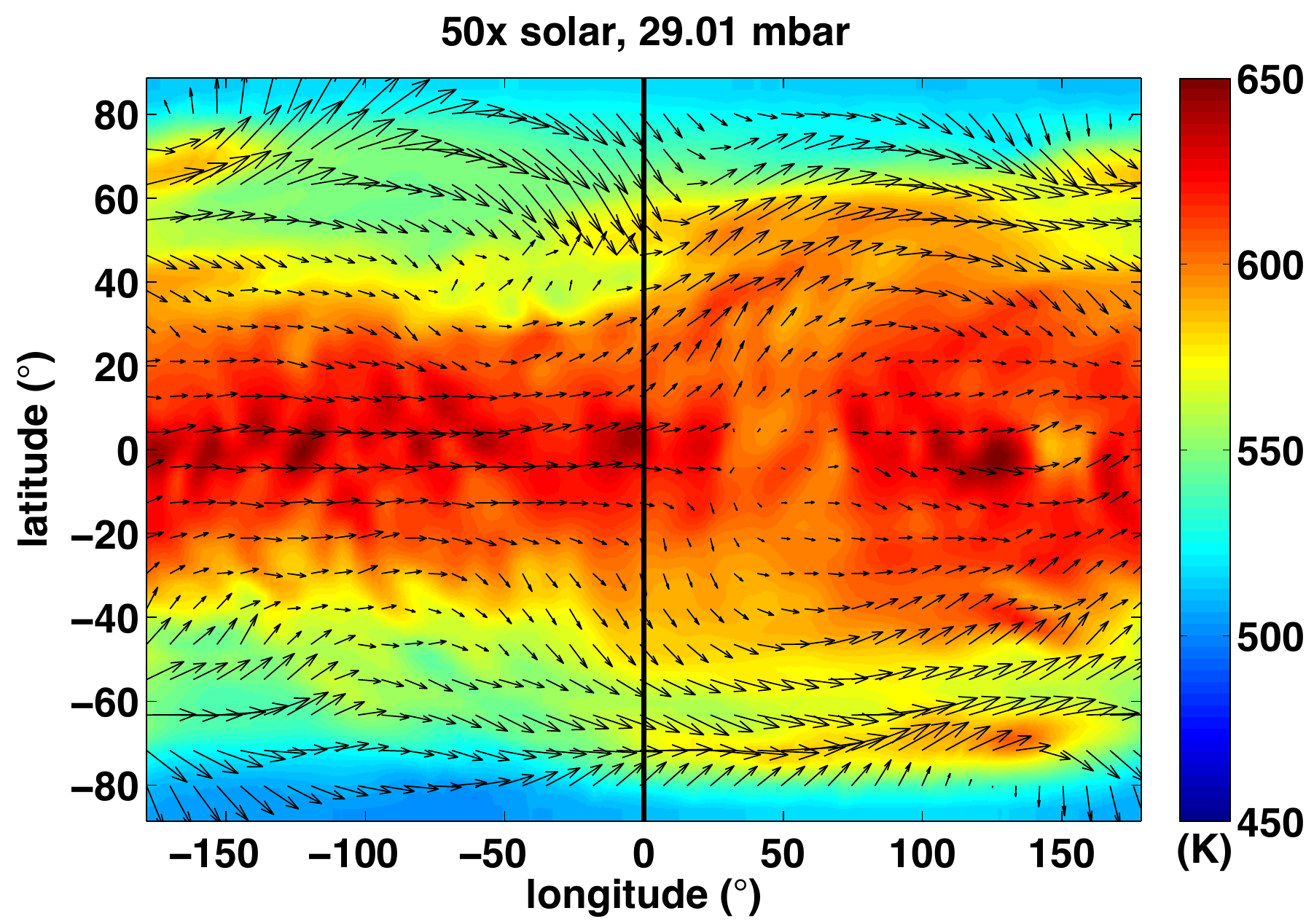}\\ 
\includegraphics[trim = 0.0in 0.0in 0.0in 0.0in, clip, width=0.32\textwidth]{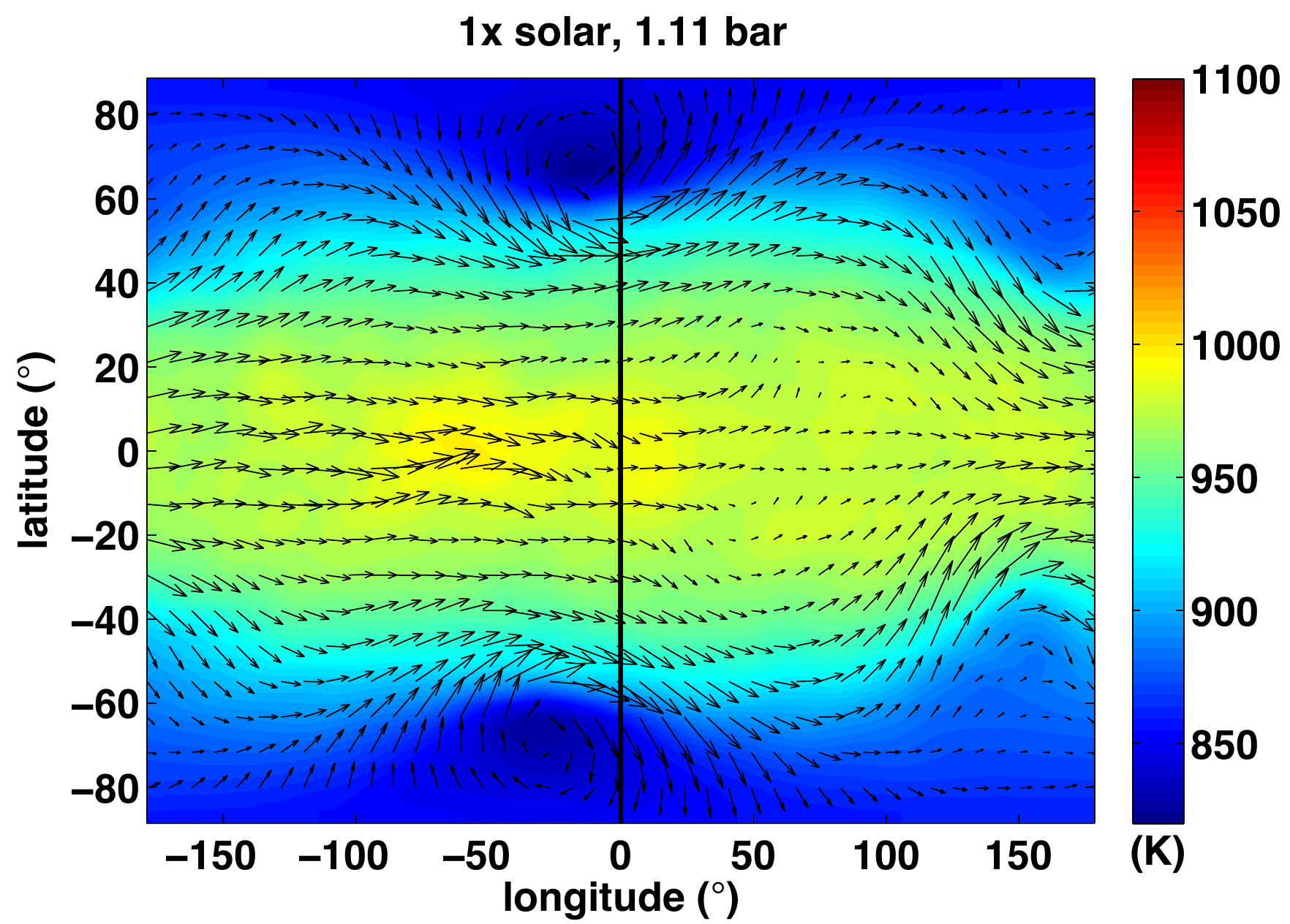} 
\includegraphics[trim = 0.0in 0.0in 0.0in 0.0in, clip, width=0.32\textwidth]{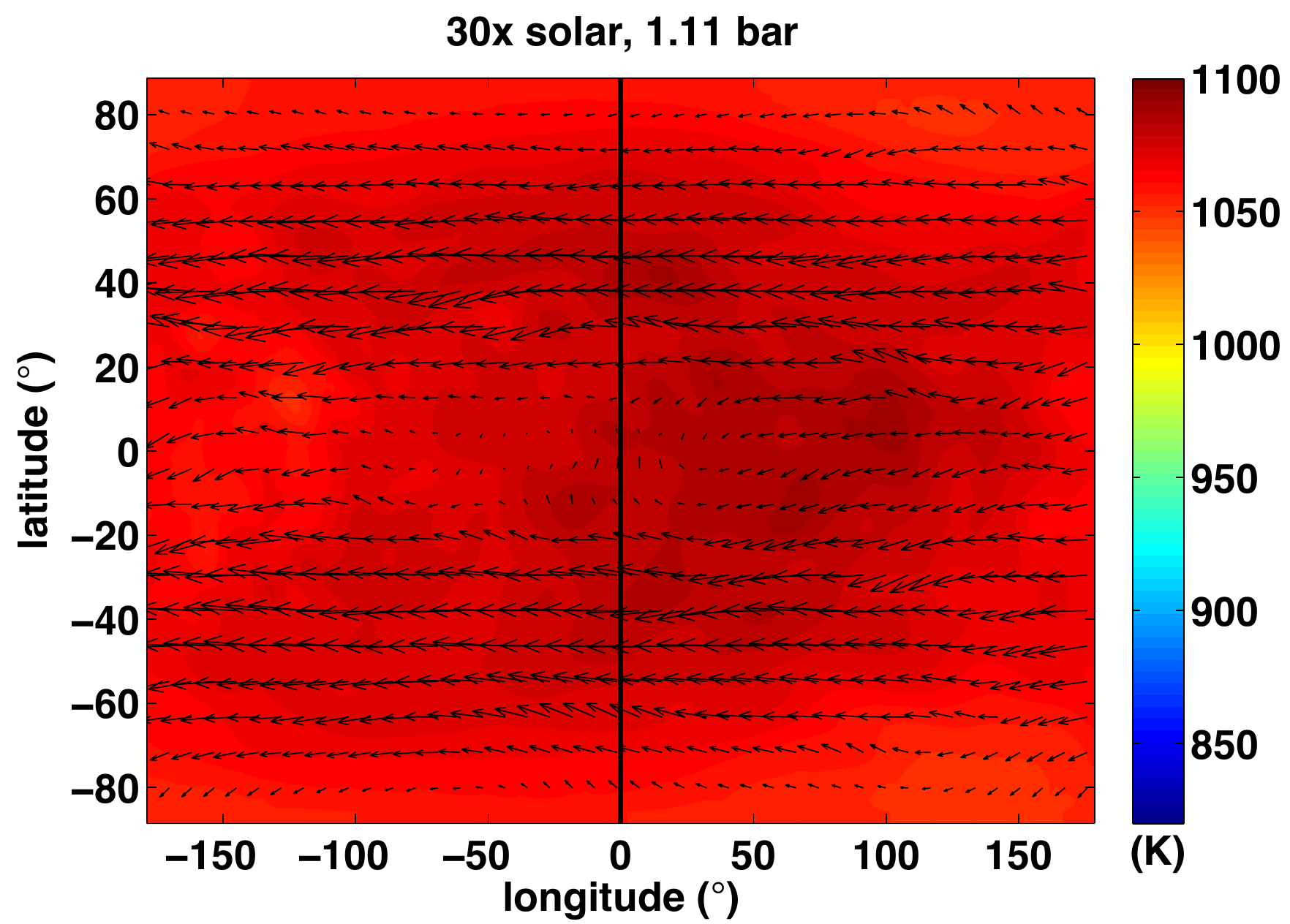} 
\includegraphics[trim = 0.0in 0.0in 0.0in 0.0in, clip, width=0.32\textwidth]{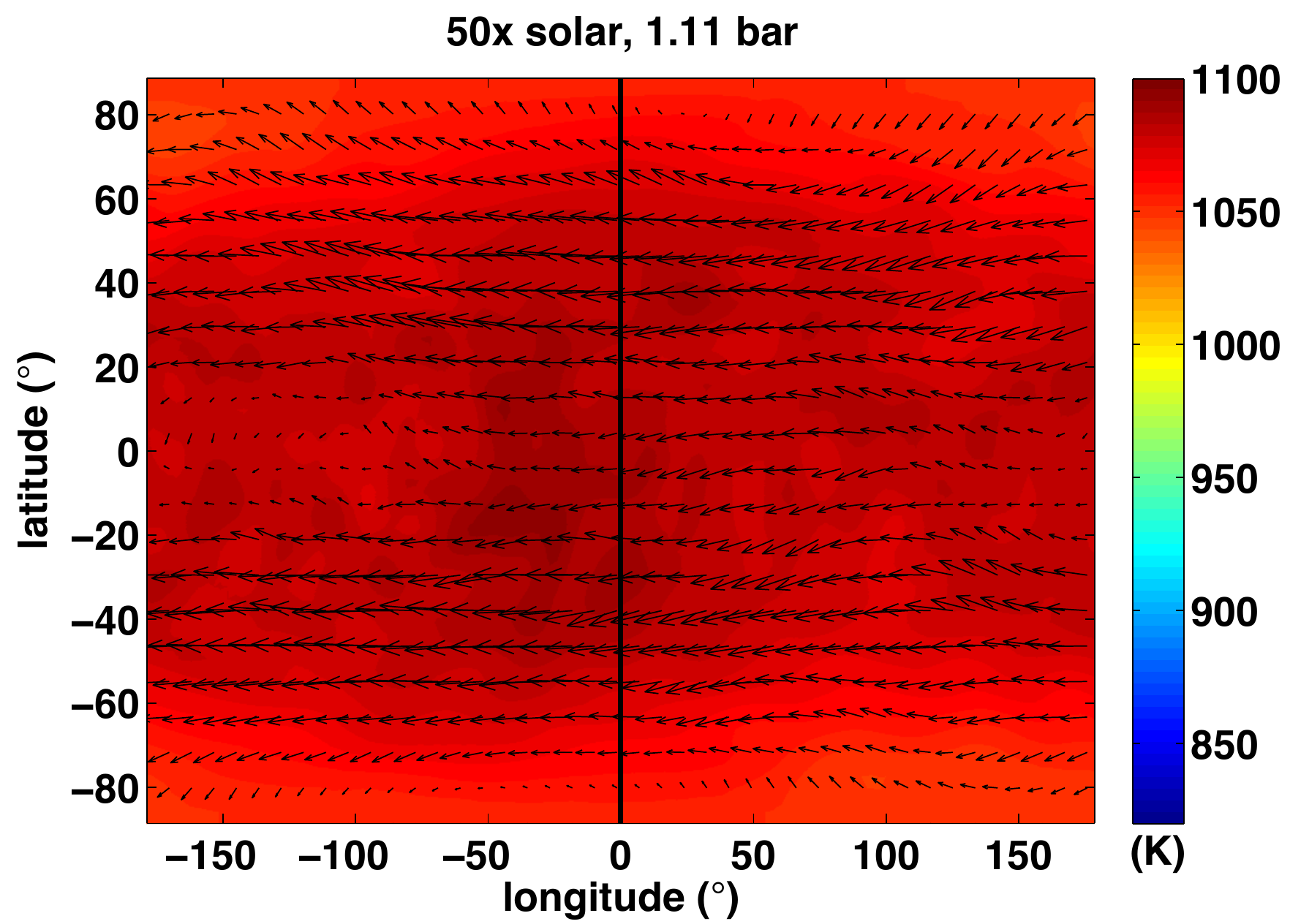}\\ 
\caption{Wind and temperature at approximately 1 mbar (top row), 30 mbar (middle row) and 1 bar (bottom row) for $\mathrm{H_2}$-dominated compositions of GJ 1214b.  Each column corresponds to atmospheric compositions of (from left to right) 1$\times$, 30$\times$, and 50$\times$ solar.  The black line denotes the substellar longitude. Each row is plotted on the same colorscale.}
\end{centering}
\end{figure*}

Two trends in circulation are seen as the metallicity is increased.  First, the peak speeds of the jet increase; equatorial jet speeds range from $\sim1.1~\mathrm{km~s^{-1}}$ in the 1$\times$ solar case to greater than 1.7 $\mathrm{km~s^{-1}}$ in the 50$\times$ solar case.  Second, the depth of the high-latitude jets decrease with increasing metallicity; jets in the solar case extend to pressures of approximately 1 bar, while the jets in higher metallicity cases extend to only $\sim$300 mbar.  Similar trends are seen in circulation models of hot Neptune GJ 436b (Lewis et al. 2010).  The trends in circulation are a result of enhanced opacity associated with higher metallicities, which leads to shallower heating in the atmosphere (Fortney et al. 2008, Dobbs-Dixon \& Lin 2008, Showman et al. 2009, Lewis et al. 2010).  

\begin{figure}
\begin{centering}
\epsscale{.80}
\includegraphics[trim = 0.0in 0.0in 0.0in 0.0in, clip, width=0.45\textwidth]{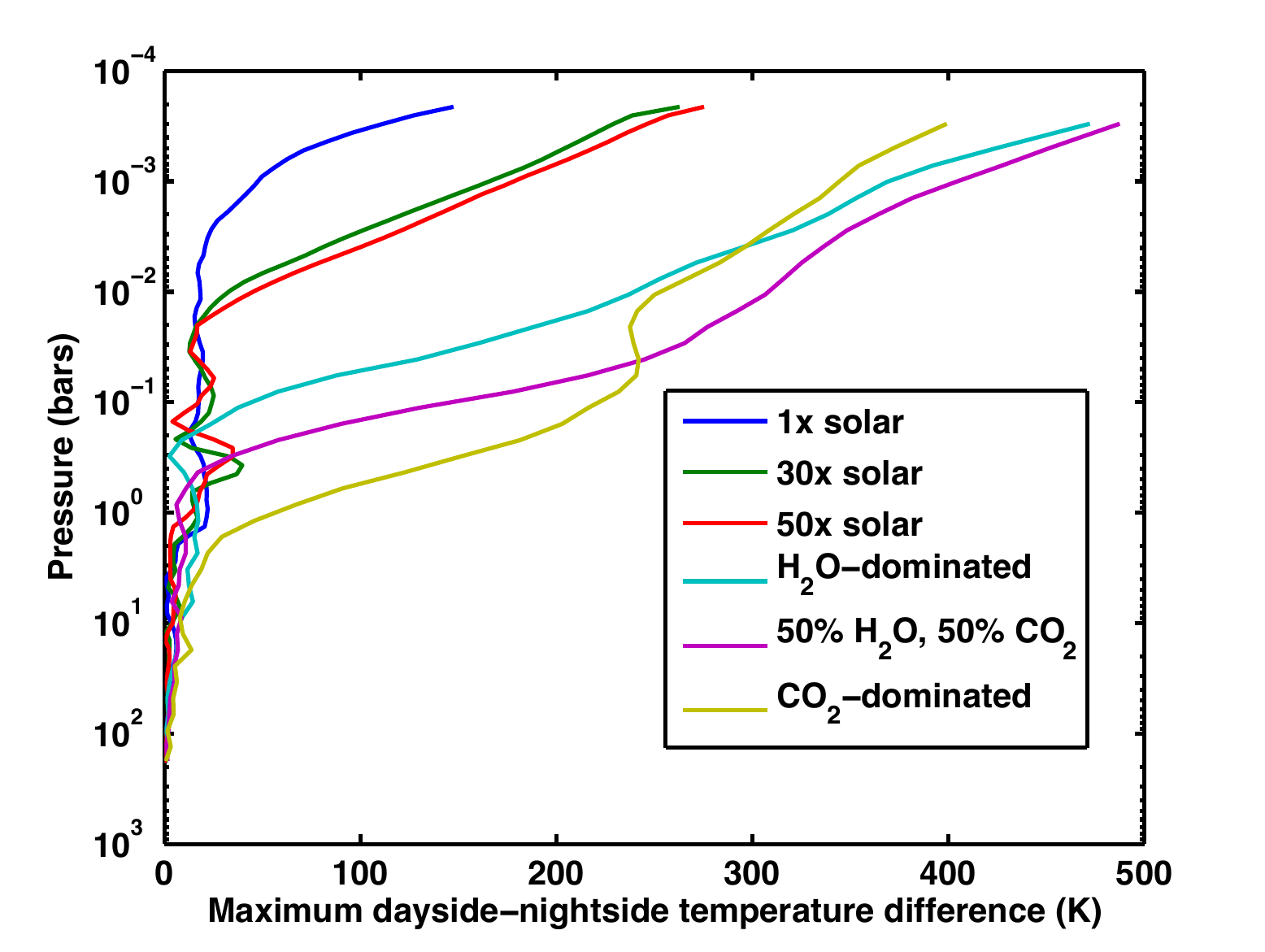}\\
\caption{Maximum day-night temperature difference as a function of pressure for all modeled atmospheric compositions of GJ 1214b. This was calculated by first computing a weighted-average of temperature as a function of longitude, then differencing the maximum and minimum temperatures on the dayside and nightside, respectively.  } 
\label{daynight_diff}
\end{centering}
\end{figure}

This enhanced opacity with higher metallicity leads to differences in the planet's horizontal and vertical temperature structure.  In our models, the temperature difference from dayside to nightside varies with height throughout observable regions of the atmosphere for all three metallicities.  However, at a given pressure, this day-night temperature difference is greater for higher metallicities.  We compare the temperature variations as a function of pressure in Figures 3 and \ref{daynight_diff}.  Figure 3 plots the wind and temperature profiles for each atmospheric metallicity at three pressure levels: 1 mbar, 30 mbar, and 1 bar, which approximately bracket the range of pressures over which infrared photons escape to space (Figure 3).  Indeed, at the shallowest pressure, 1 mbar, the 50$\times$ solar model exhibits the highest day-night temperature variations.  At 30 mbar, day-night temperature differences are small, but the 50$\times$ solar case nevertheless exhibits the largest temperature variation from equator to pole.  At 1 bar, only the 1$\times$ solar case exhibits significant temperature variation, as stellar energy is deposited deeper at low metallicity.  These trends are illustrated further in Figure \ref{daynight_diff}, which plots the maximum dayside-nightside temperature difference at each pressure level for each atmospheric composition.  This is calculated at each pressure level by first latitudinally-weighting the temperature at each longitudinal slice.  We then determine whether each slice is on the dayside or nightside, then subtract the minimum (weighted) temperature on the nightside from the maximum (weighted) temperature on the dayside to determine the maximum dayside-nightside temperature difference.  As shown in Figure 4, above photospheric pressures (less than $\sim$10 mbar), the day-night temperature variation at each pressure increases with increasing metallicity.   Given the expectation that the day-night heating drives the equatorial superrotation \citep{sp2011,kataria2013}, this would imply stronger superrotation with increasing metallicity, qualitatively explaining the trend seen in Figure 2.  At pressures greater than 10 mbar, where radiative time constants are longer, the temperature varies in longitude by less than $\sim$25 K.   These trends in temperature and wind structure will affect resultant synthetic lightcurves and spectra (see Section 5).  

\begin{figure*}
\begin{centering}
\epsscale{.80}
\includegraphics[trim = 0.0in 0.0in 0.0in 0.0in, clip, width=0.43\textwidth]{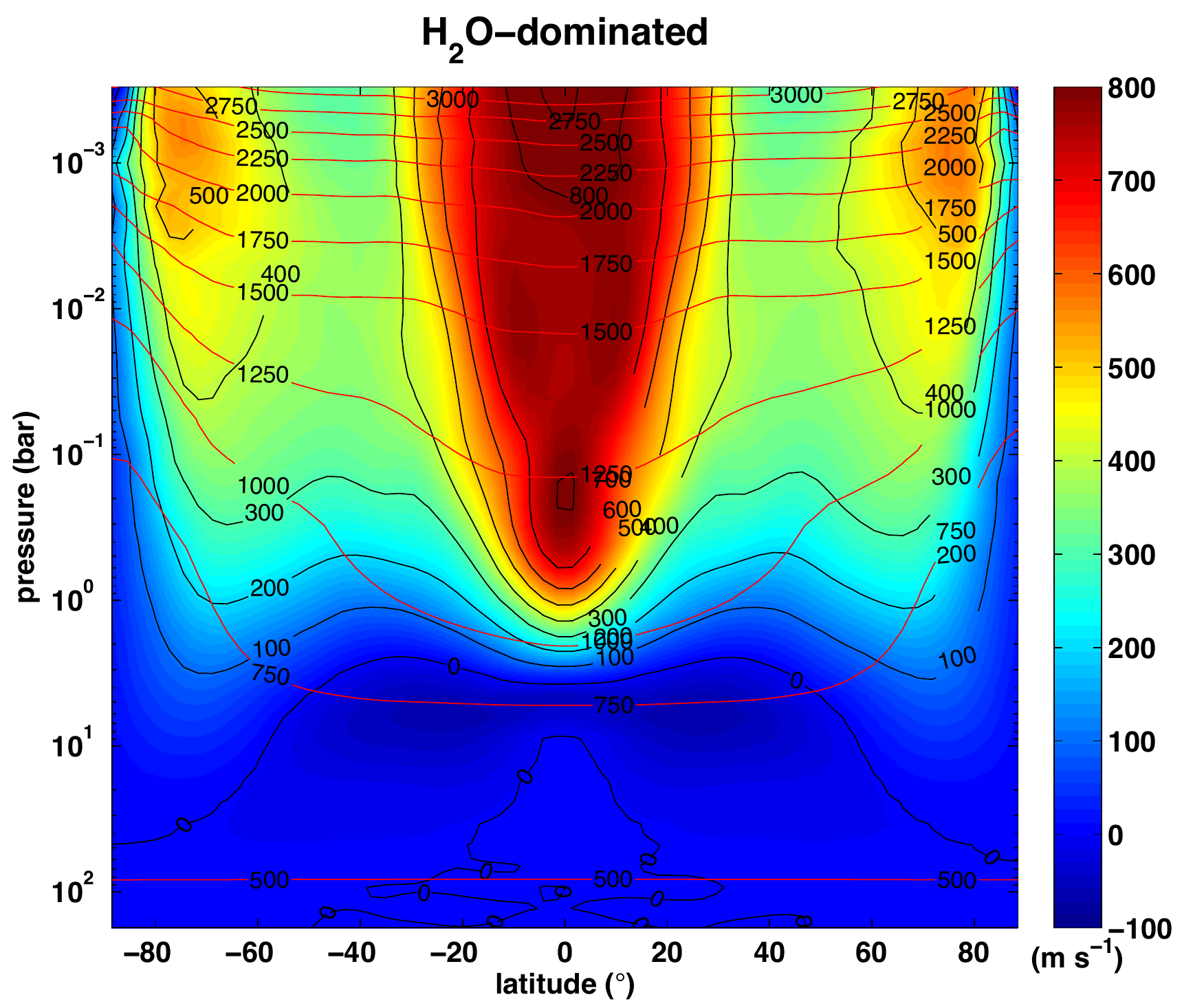}
\includegraphics[trim = 0.0in 0.0in 0.0in 0.0in, clip, width=0.48\textwidth]{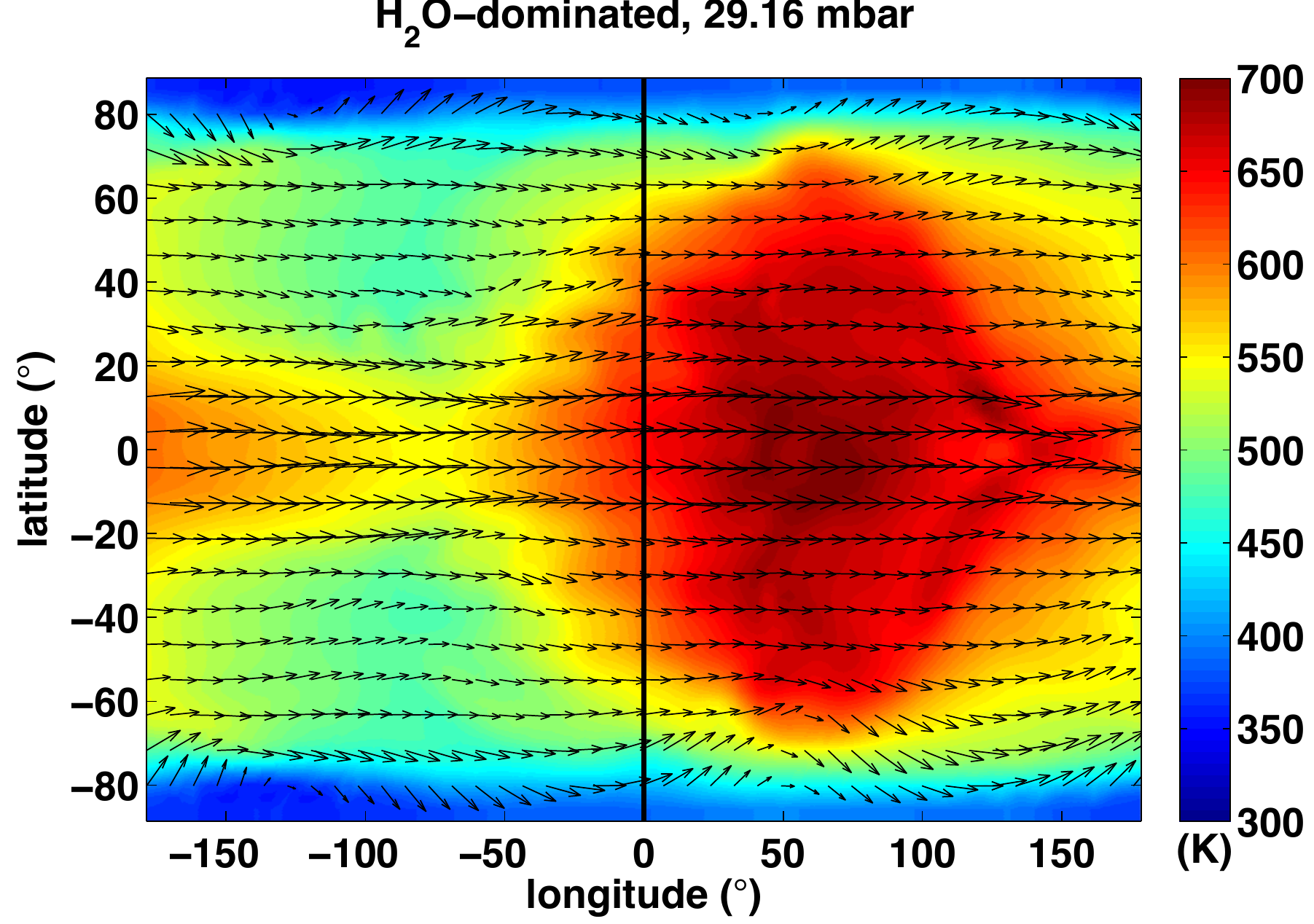}\\
\includegraphics[trim = 0.0in 0.0in 0.0in 0.0in, clip, width=0.43\textwidth]{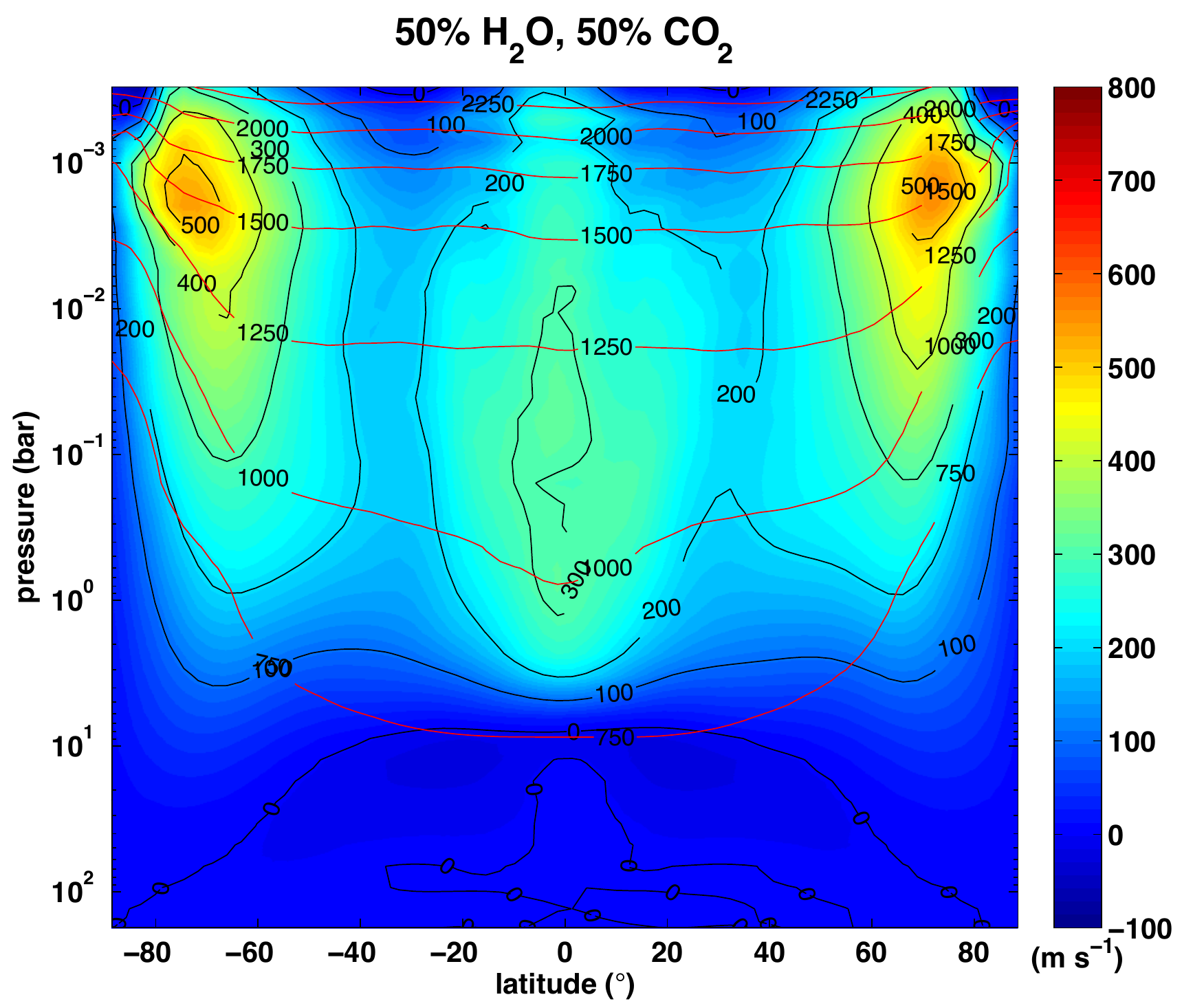}
\includegraphics[trim = 0.0in 0.0in 0.0in 0.0in, clip, width=0.48\textwidth]{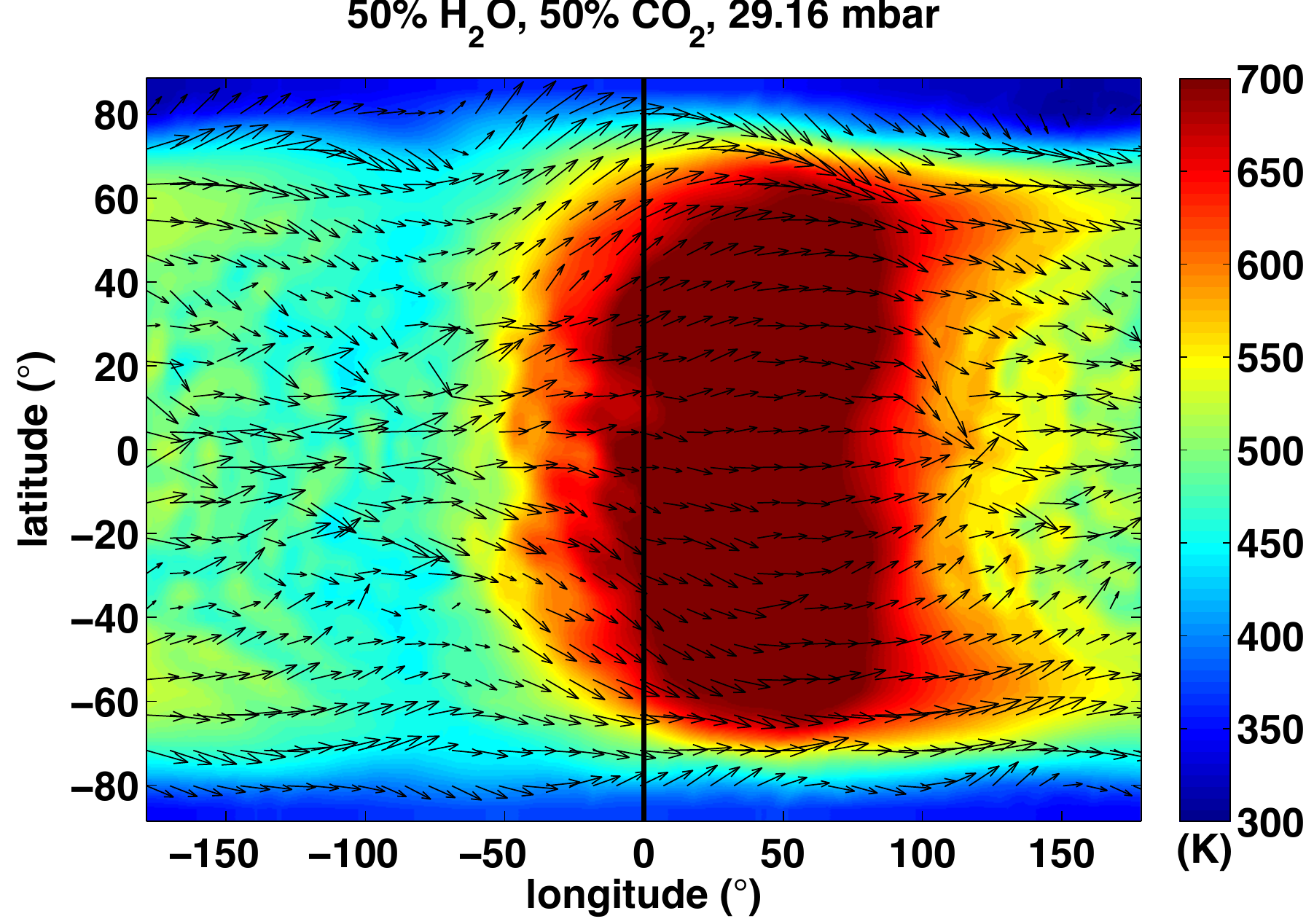}\\
\includegraphics[trim = 0.0in 0.0in 0.0in 0.0in, clip, width=0.43\textwidth]{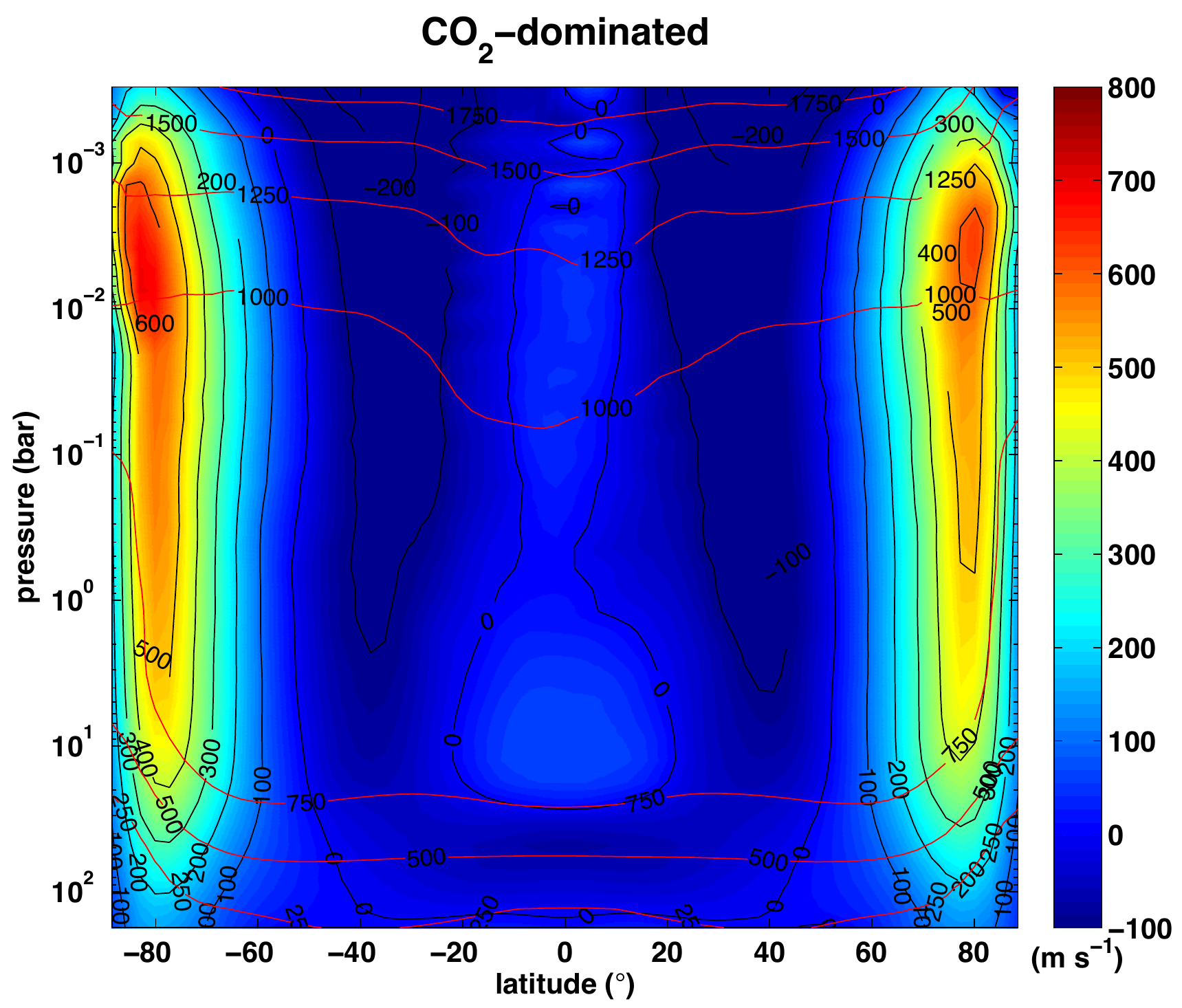}
\includegraphics[trim = 0.0in 0.0in 0.0in 0.0in, clip, width=0.48\textwidth]{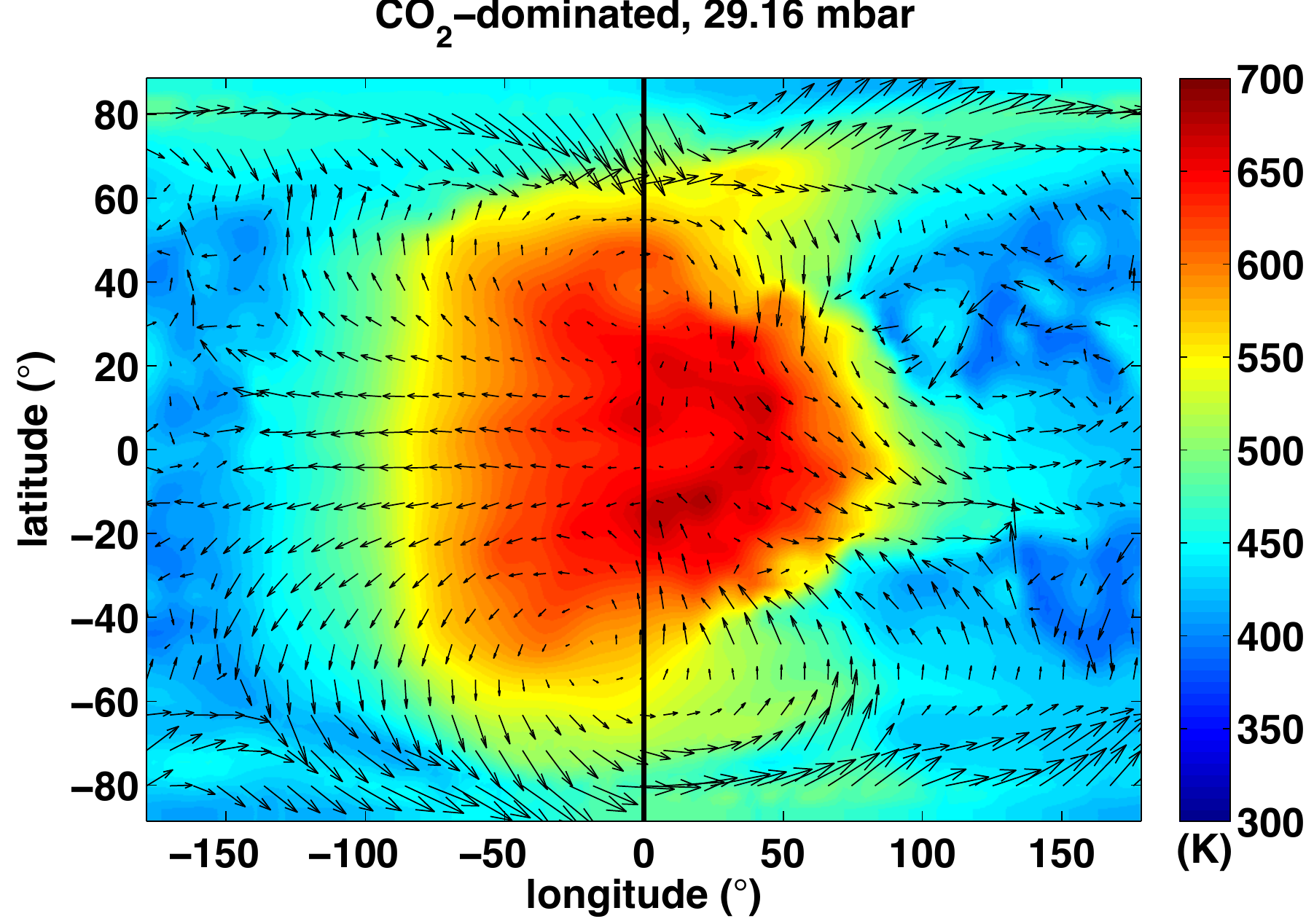}\\
\caption{Zonal-mean zonal wind (left column) and wind and temperature at 30 mbar (right column) for high MMW atmospheric compositions of GJ 1214b.  Each pair of plots correspond to atmospheric compositions of (from top to bottom) 99\% $\mathrm{H_2O}$, 1\% $\mathrm{CO_2}$; 50\% $\mathrm{H_2O}$, 50\% $\mathrm{CO_2}$; and 99\% $\mathrm{CO_2}$, 1\% $\mathrm{H_2O}$.  Zonal-mean isentropes are overplotted in red in intervals of 250 K.  The panels in each column are shown with the same colorscale.}
\label{highmmw_plots}
\end{centering}
\end{figure*}

\begin{figure*}
\begin{centering}
\includegraphics[trim = 0.5in 2.6in 1.0in 3.0in, clip, width=0.43\textwidth]{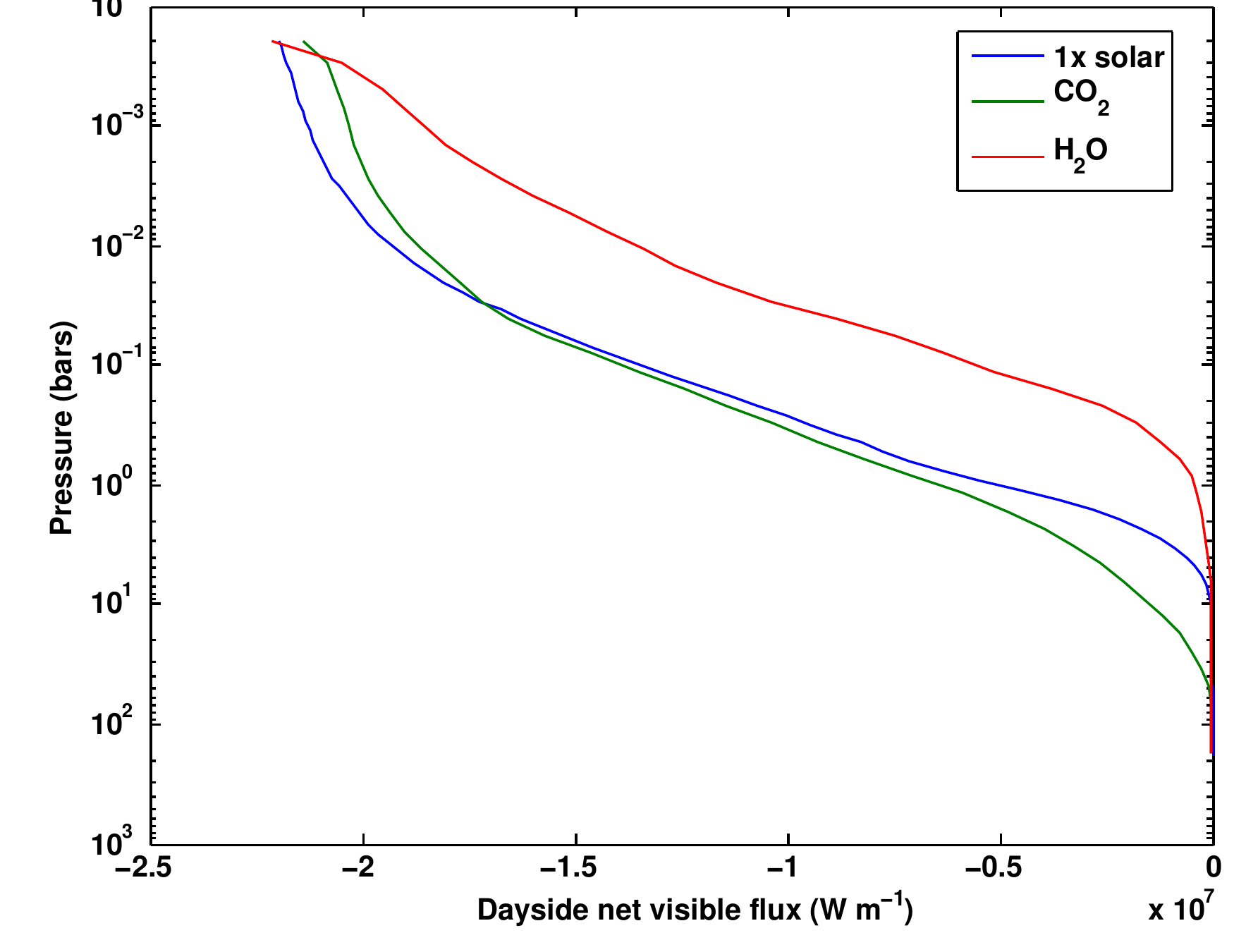}
\includegraphics[trim = 0.5in 2.6in 1.0in 3.0in, clip, width=0.43\textwidth]{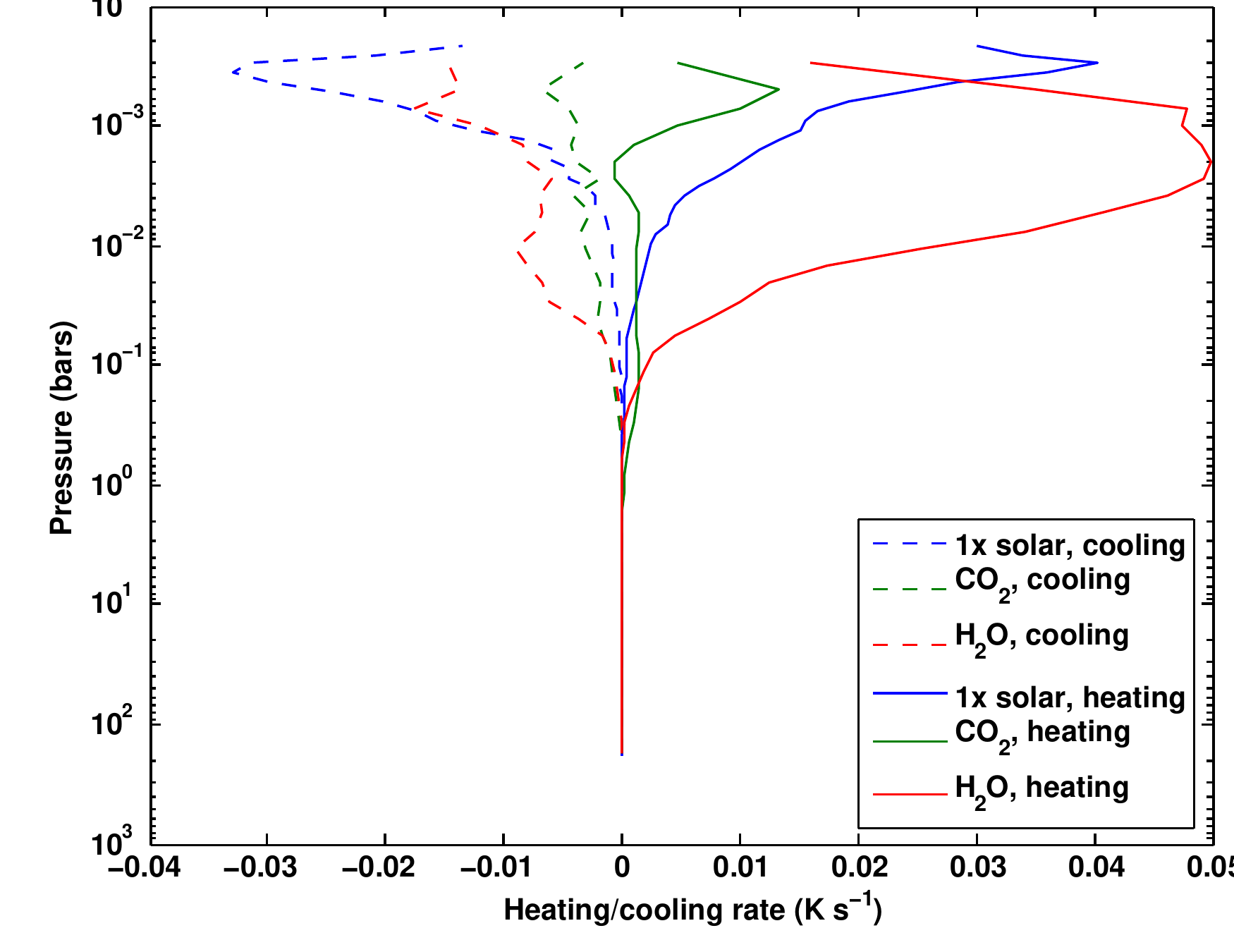}\\
\includegraphics[trim = 0.5in 2.6in 1.0in 3.0in, clip, width=0.43\textwidth]{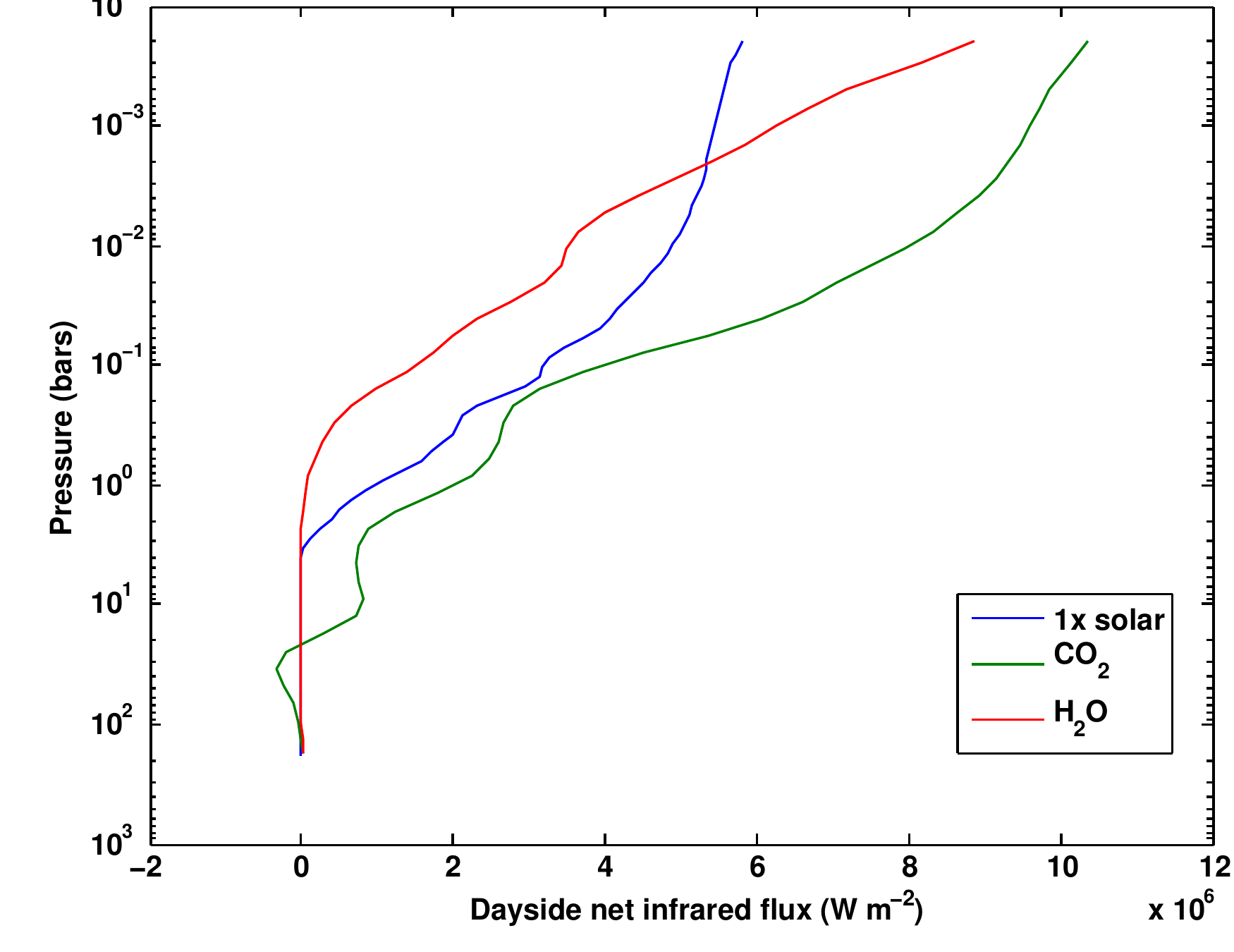}
\includegraphics[trim = 0.5in 2.6in 1.0in 3.0in, clip, width=0.43\textwidth]{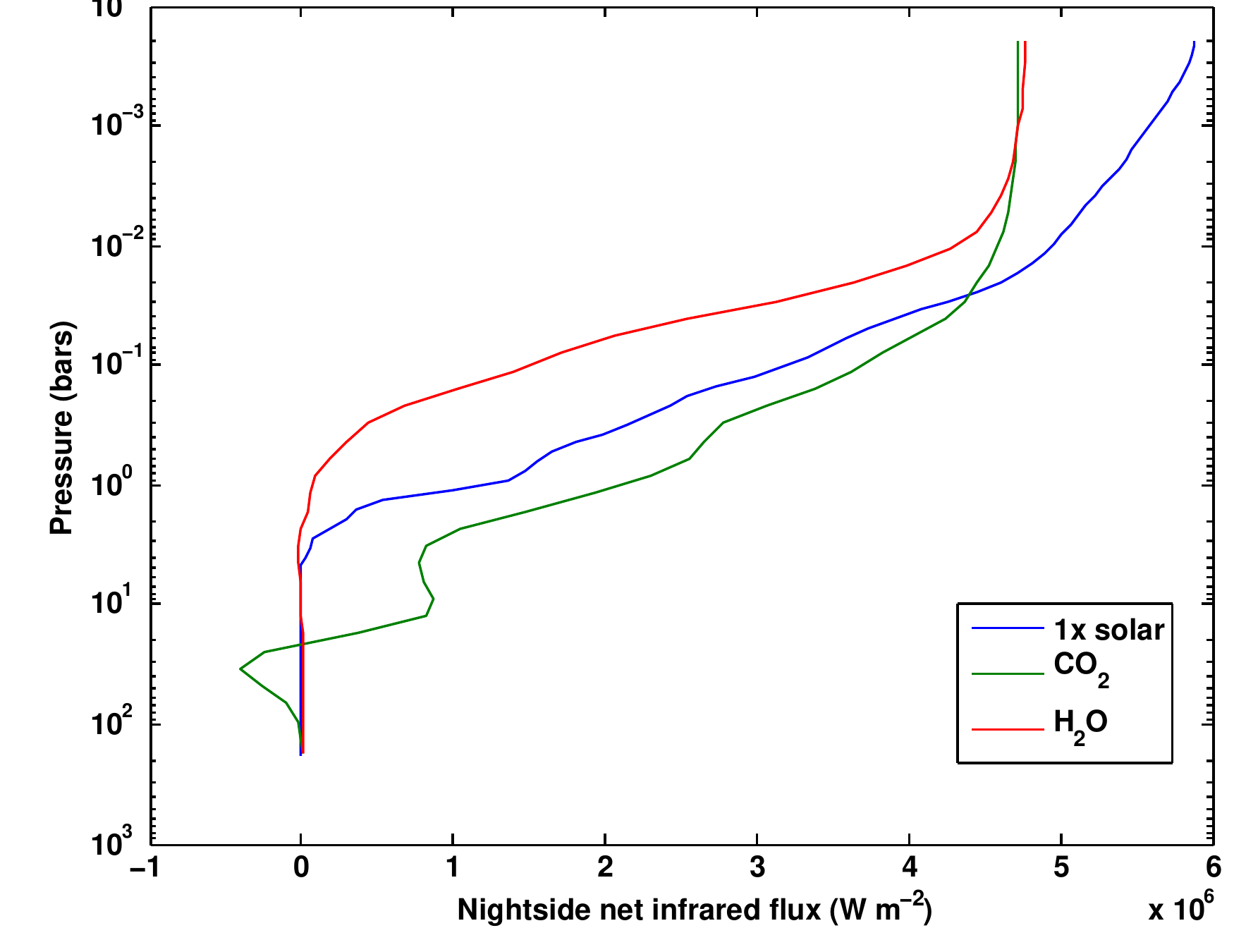}
\caption{Dayside net visible flux (top left), heating/cooling rates (top right), dayside net infrared flux (bottom left) and nightside net infrared flux (bottom right) as a function of pressure, for each major atmospheric composition: H-dominated (1$\times$ solar composition, blue), $\mathrm{H_2O}$-dominated (red) and $\mathrm{CO_2}$-dominated (green).  Fluxes are in units of $\mathrm{W~m^{-2}}$, while heating/cooling rates are in units of $\mathrm{K~s^{-1}}$. }
\label{flux_plots}
\end{centering}
\end{figure*}

\begin{figure}
\begin{centering}
\includegraphics[width=0.5\textwidth]{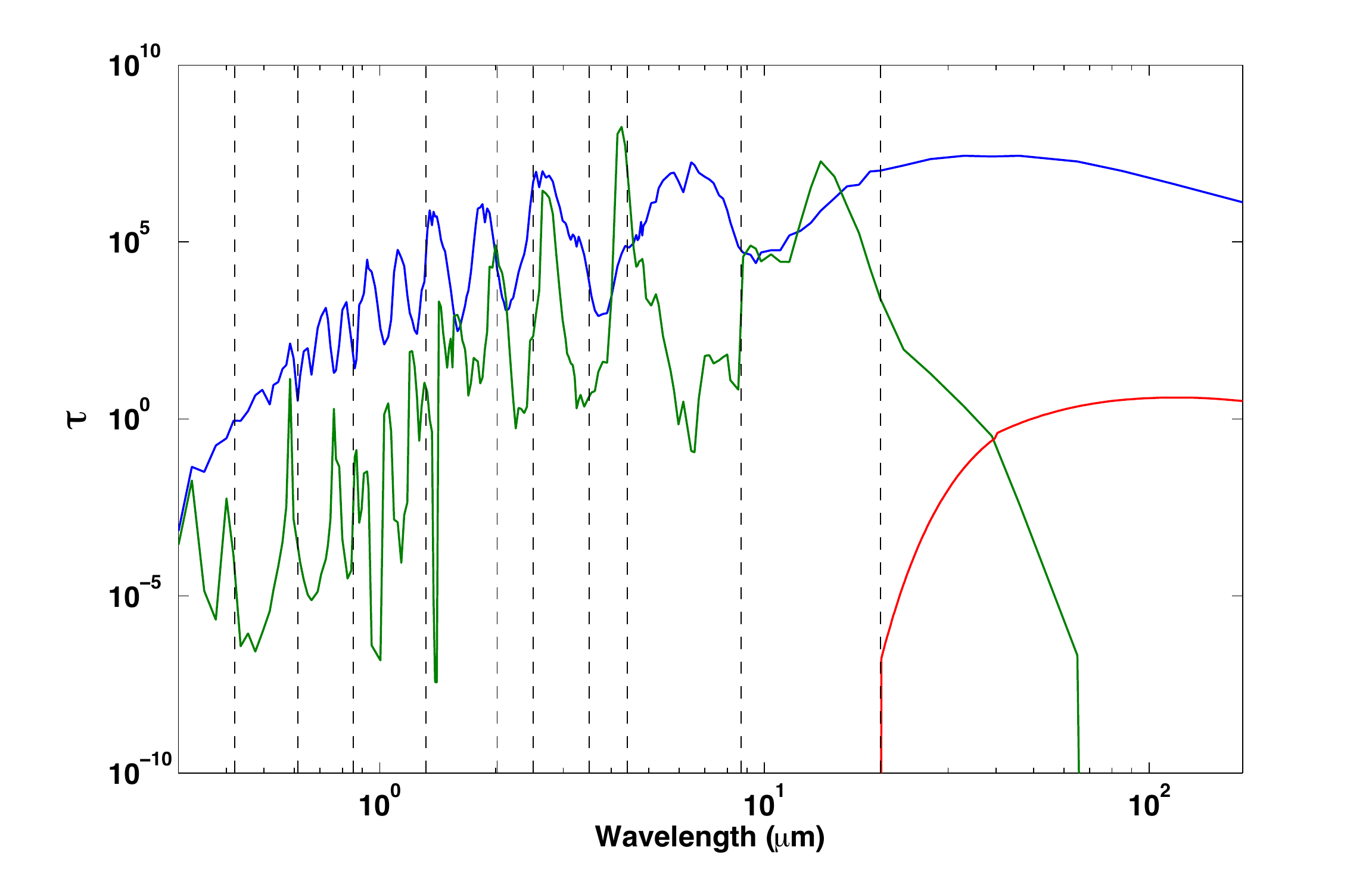}
\caption{Optical depth as a function of wavelength for pure-water (blue), pure-$\mathrm{CO_2}$ (green), and pressure-induced absorption (PIA) due to $\mathrm{CO_2}$-$\mathrm{CO_2}$ collisions (red).  The dotted grey lines denote the boundaries of the 11 spectral bins used in the correlated-$k$ calculation.  Note that the $\mathrm{CO_2}$-$\mathrm{CO_2}$ PIA is only prominent in the longest wavelength (shortest wavenumber) bin. }
\label{taus}
\end{centering}
\end{figure}

\subsection{Water and Carbon-dioxide atmospheric compositions}
A comparison of high-MMW atmospheric compositions yield major differences in the dynamical and temperature regimes of GJ 1214b. If the atmosphere is $\mathrm{H_2O}$-dominated (Figure \ref{highmmw_plots}, top row), the atmosphere still possesses an equatorial superrotating jet, with peak speeds of $\sim900~\mathrm{m~s^{-1}}$, and high-latitude jets with speeds exceeding 500 $\mathrm{m~s^{-1}}$.  For a $\mathrm{CO_2}$-dominated atmosphere, however, equatorial superrotation is much weaker; instead, the dynamics are dominated by high-latitude jets, with peak speeds exceeding 500 $\mathrm{m~s^{-1}}$ (Figure \ref{highmmw_plots}, bottom row).  The $\mathrm{50\%~CO_2}$, $\mathrm{50 \%~H_2 O}$ case, as expected, exhibits an intermediate behavior, whereby the atmosphere is dominated by broad, high-latitude jets and moderate equatorial superrotation (Figure \ref{highmmw_plots}, middle row).  However, all three cases have higher equator-to-pole and day-night temperature variations at photospheric pressures ($\sim$10 mbar) than the low-MMW models.  These $>$100 K variations extend as deep as 100 mbar, an order of magnitude greater than the low-MMW cases (Figure \ref{daynight_diff}).  

The changes in dynamical and temperature regimes between low- and high-MMW atmospheres and between water- and $\mathrm{CO_2}$-dominated atmospheres can be attributed to differences in the vertical opacity structure and hence heating budget.  For a $\mathrm{CO_2}$-dominated atmosphere, the atmosphere is more transparent to visible radiation.  Hence, the stellar energy is deposited deeper in the atmosphere as compared to hydrogen- and water-dominated atmospheres.  The qualitative picture can be further confirmed by plotting the heating/cooling rates and visible and IR fluxes on the dayside and nightside (Figure \ref{flux_plots}).  The top left panel shows the dayside net visible flux, which has a net downward direction.  The water case absorbs the incoming stellar energy much higher in the atmosphere as compared to the solar and $\mathrm{CO_2}$ cases, which corresponds to a much larger specific heating rate at the top of the atmosphere where the atmospheric mass is much less (top right panel).  Note also that the heating and cooling rates are smallest for the $\mathrm{CO_2}$-dominated case, helping to explain the weak superrotation.  The large variation in visible flux with height for the water-dominated case leads to a large specific heating rate at low pressures, where the atmospheric mass is less.  The bottom two panels plot the net IR flux at the substellar and antistellar points, respectively.  They show that the water-dominated case also emits flux at lower pressures compared to the other two compositions.  Overall, the plots show that the $\mathrm{CO_2}$-dominated case absorbs energy deepest, and the water-dominated case highest.  

Based on the results presented in \cite{sp2011}, one would expect that the 1$\times$ solar case, which has the strongest superrotation, should absorb visible flux at lower pressures compared to the other two compositions, where day-night temperature variations and forcing are largest.  However, as described above, the water-dominated case has the shallowest flux deposition.  This suggests that the differences in specific heat (and therefore scale height) might also play a role in the differences in energy budget and dynamical regimes.  To test this hypothesis, we ran two models, the first with 1$\times$ solar atmospheric opacities but a specific heat, mean-molecular weight, and scale height set to the $\mathrm{CO_2}$-dominated value, and a second model which has the reverse ($\mathrm{CO_2}$-dominated atmospheric opacities, 1$\times$ solar specific heat, MMW and scale height).  The 1$\times$ solar opacity case does show flow features similar to that of the $\mathrm{CO_2}$-dominated case in Figure \ref{highmmw_plots}, with high latitude jets and weak superrotation at the equator.  A detailed analysis of these differences, specifically for the $\mathrm{CO_2}$-dominated case, will be a task for future studies.

\section{Comparison to other circulation models of GJ 1214b}
We can compare our results to the other circulation models of GJ 1214b, particularly Menou (2012), which models three of the atmospheric compositions included in this paper (water-dominated, 1$\times$ and 30$\times$ solar), though with a different circulation model (the Intermediate General Circulation Model), radiative transfer scheme (double-grey), and model setup.  In comparing the hydrogen-dominated models (see Figure 2 in Menou 2012), one can see broad agreement, with equatorial superrotation in the 1-2 $\mathrm{km~s^{-1}}$ range, and high-latitude eastward winds.   However, the jet structure is different--the equatorial jets in Menou (2012) extend to deeper pressures than our models.  These differences could stem from differences in radiative transfer scheme, but also differences in bottom boundary (10 bars vs. 100 bars in our model).  

In comparing water-dominated circulation models of GJ 1214b, we can also include results from Zalucha et al. (2013), who use a different setup of the MITgcm coupled to a Newtonian relaxation scheme, including a surface at varying pressures.  In all three models there is again broad agreement, with an eastward equatorial jet with a width of approximately 50-60 degrees.  However, the models again differ in jet speeds and structure.  Equatorial wind speeds are greatest in Menou (2012), and Zalucha et al. (2013) model the weakest.  Furthermore, both Menou (2012) and our results include eastward winds at high latitudes, while Zalucha et al. has westward winds at the same latitudes.  These differences are most likely due to differences in the bottom boundary and radiative heating schemes.  

\begin{figure*}
\begin{centering}
\epsscale{.80}
\includegraphics[trim = 0.0in 0.0in 0.0in 0.0in, clip, width=0.43\textwidth]{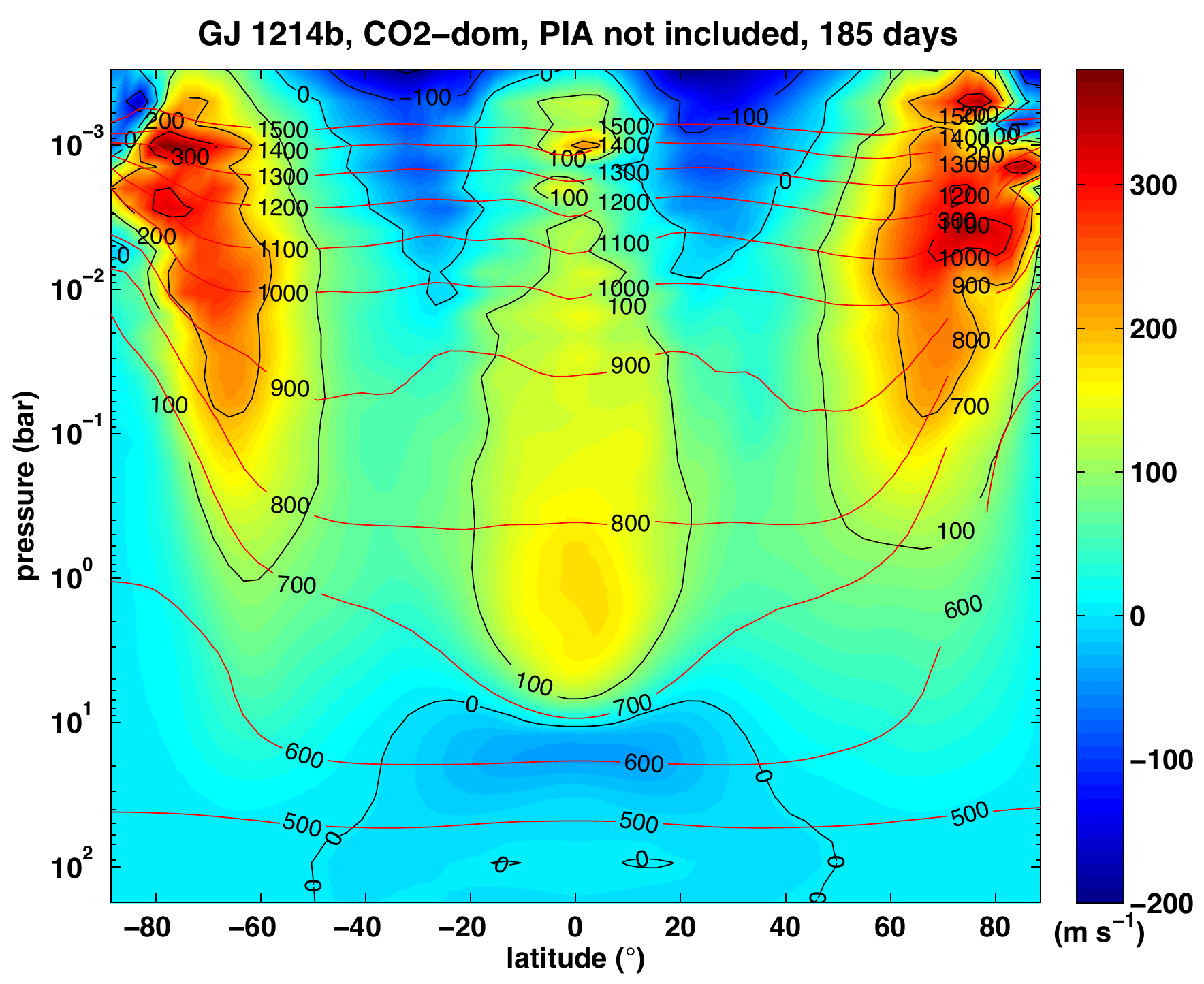}
\includegraphics[trim = 0.0in 0.0in 0.0in 0.0in, clip, width=0.43\textwidth]{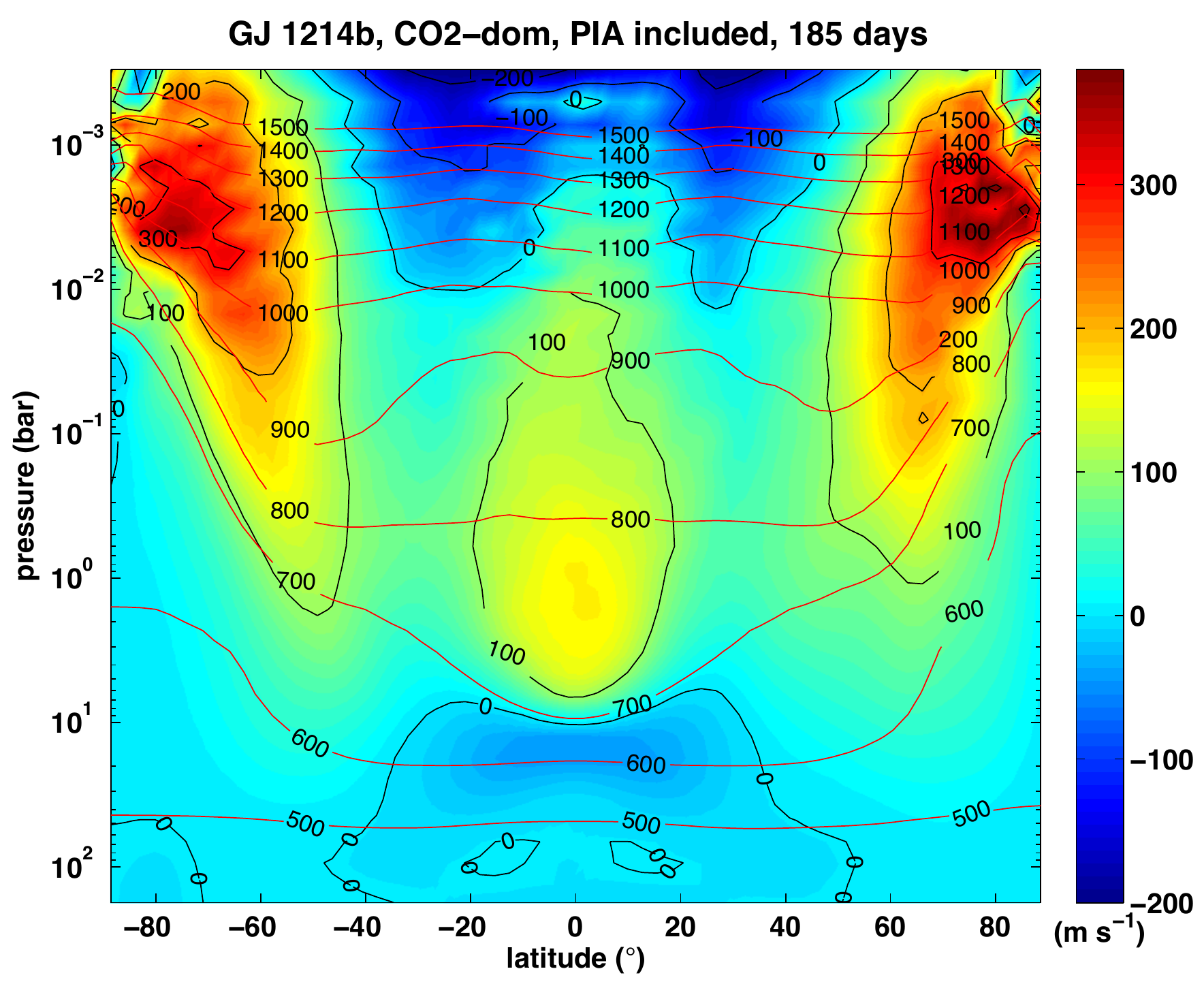}\\
\includegraphics[trim = 0.0in 0.0in 0.0in 0.0in, clip, width=0.43\textwidth]{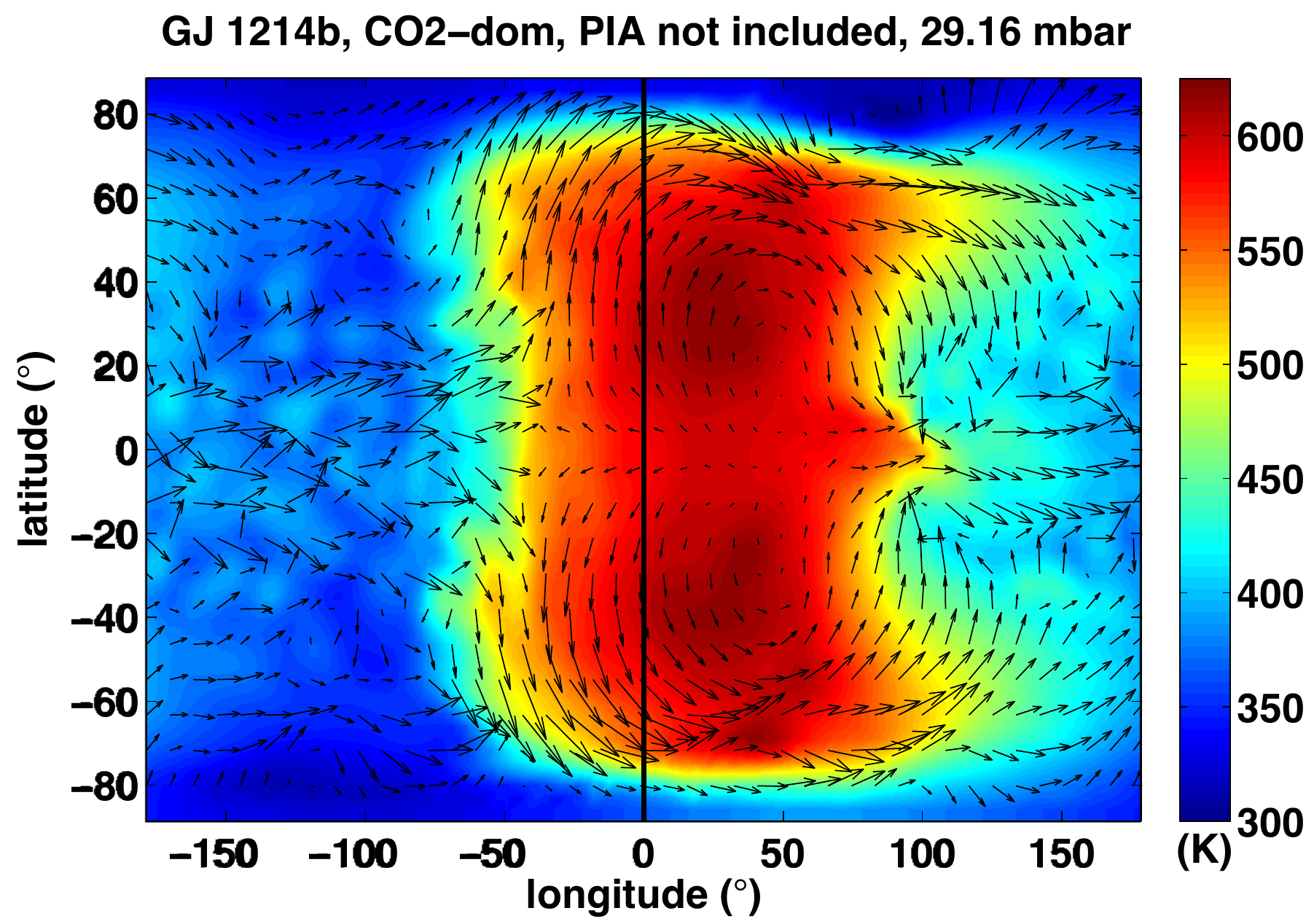}
\includegraphics[trim = 0.0in 0.0in 0.0in 0.0in, clip, width=0.43\textwidth]{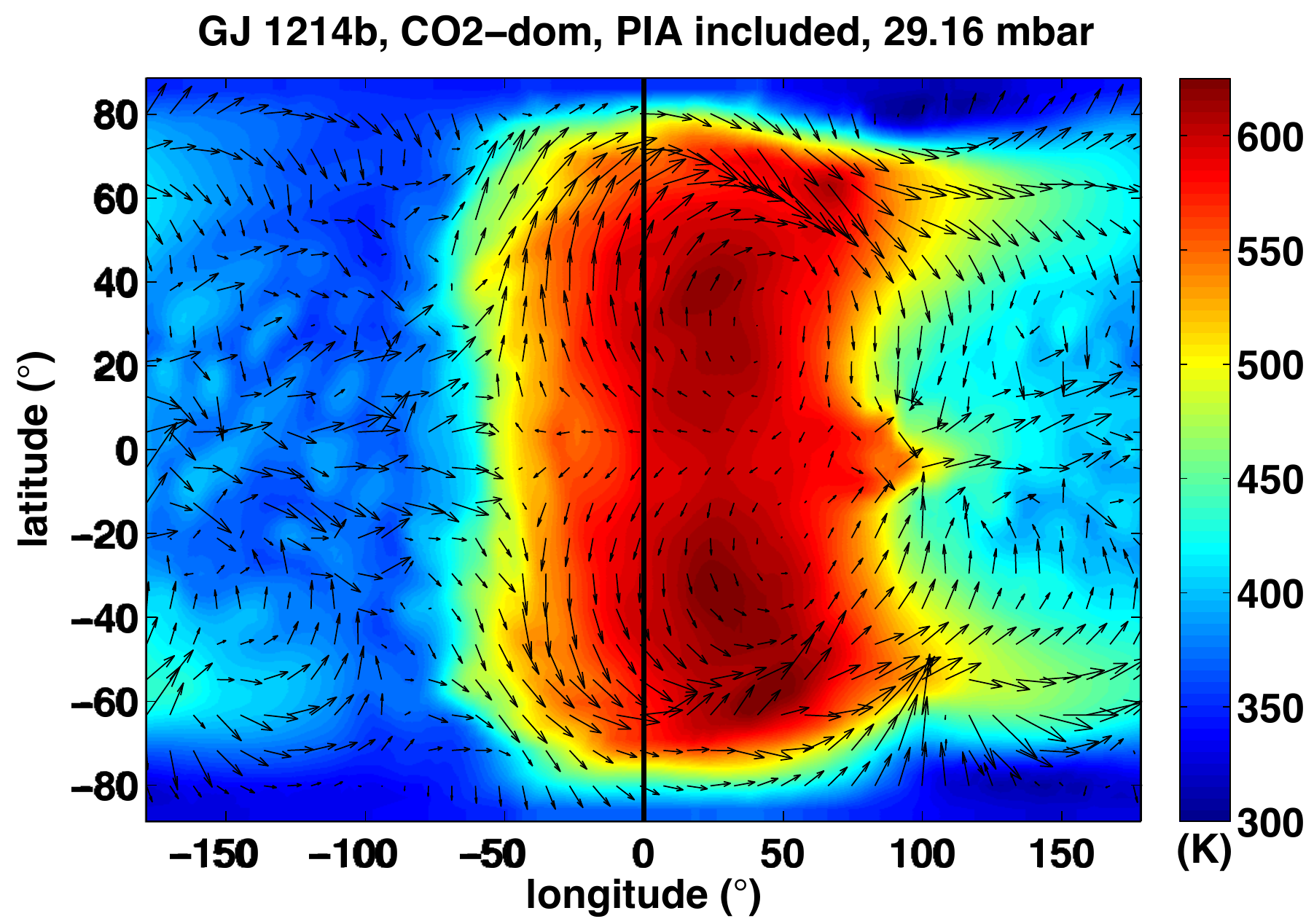}\\
\caption{Snapshots of the zonal-mean zonal wind (top row) and wind and temperature profiles at 30 mbar (bottom row) during the spin-up phase for $\mathrm{CO_2}$-dominated atmospheric compositions with (right column) and without (left column) the inclusion of opacity effects due to pressure-induced absorption (PIA) from $\mathrm{CO_2}$-$\mathrm{CO_2}$ collisions.  Each pair of plots are shown on the same colorscale.  Zonal-mean isentropes are plotted in red, while the black vertical lines in both wind/temperature profiles indicate the longitude of the substellar point.}
\label{co2pia_plots}
\end{centering}
\end{figure*}

\section{Including effects due to $\mathrm{CO_2}$-$\mathrm{CO_2}$ PIA}
While we already include opacity effects due to pressure-induced absorption (PIA) from $\mathrm{H_2}$-$\mathrm{H_2}$ and $\mathrm{H_2}$-$\mathrm{He}$ collisions (see Kataria et al. 2013), here we test the importance of including opacity effects due to PIA from $\mathrm{CO_2}$-$\mathrm{CO_2}$ collisions. Figure \ref{taus} compares the optical depth, $\tau$, of this opacity source with the optical depths for a pure water and pure $\mathrm{CO_2}$ atmosphere.  This optical depth is calculated as a product of the number of molecules per $\mathrm{cm^{-2}}$, $\mathcal{N}$, and the absorption coefficients, $\mathcal{K}$.  The value of $\mathcal{N}$ is defined as $nH$, where $n$ is the number density (in units of $\mathrm{m^{-3}}$) and $H$ is the scale height.  This can be further simplified using the ideal gas law as $\mathcal{N}={P/mg}$.  Here we calculate the optical depths for each composition in each wavelength interval at a temperature of 725 K and a pressure of 1 bar ($\mathrm{10^{6}}$ cgs).

As shown in Figure \ref{taus}, the $\mathrm{CO_2}$-$\mathrm{CO_2}$ PIA is most important in the longest wavelength (shortest wavenumber) frequency bin (denoted by grey dotted lines).  Only a small fraction of the planet's flux is emitted in this wavelength range, and therefore we expect that the inclusion of $\mathrm{CO_2}$-$\mathrm{CO_2}$ PIA should not significantly affect the dynamical structure.  Figure \ref{co2pia_plots} compares the transient spin-up phase of two $\mathrm{CO_2}$-dominated runs with (right column) and without (left column) the inclusion of $\mathrm{CO_2}$-$\mathrm{CO_2}$ PIA in zonal-mean zonal wind (top row) and wind/temperature profiles at 30 mbar (bottom row).  There are minor differences between both cases; the westward flow at the top of the atmosphere extends to deeper pressures at the equator when PIA is included, and the PIA case exhibits a slightly different flow pattern at $\sim$30 mbar.  However, the bulk features remain the same: the speeds and horizontal/vertical extent of the high latitude jets, and the temperature and shape of the hottest regions on the dayside do not differ significantly.  Therefore, while it is important to include this opacity source, it does not dramatically change the dynamical and thermal structure of the atmosphere.

\section{Simulated lightcurves and spectra}
Using the outputs from our model integrations, we can generate lightcurves and spectra of GJ 1214b for each atmospheric composition.  Most ground- and space-based observations of GJ 1214b have been obtained during transit, but their flat transmission spectra suggest the presence of clouds that prevent easy characterization of the atmosphere.  Therefore, only dayside emergent flux spectra obtained at secondary eclipse and lightcurves will be able to constrain the planet's atmospheric composition.  Observations with the {\it Spitzer Space Telescope} were able to detect secondary eclipse (Fraine et al. 2013, Gillon et al. 2013) and future instrumentation on the { \it James Webb Space Telescope} (JWST) and the {\it Thirty-Meter Telescope} (TMT) will improve on those observations.  In anticipation of these and other future instruments, we generate theoretical spectra and lightcurves at wavelength bands not specific to any particular instrument (see below).  In this way, observers may use these theoretical lightcurves and spectra to select the wavelengths that best suit their efforts.  

We choose to focus on atmospheric compositions of 50$\times$ solar and 99\% $\mathrm{H_2O}$/1\% $\mathrm{CO_2}$ (water-dominated), as these two models best illustrate the differences in emergent flux spectra and lightcurves that arise from differences in circulation and temperature structures.  As discussed in Sections 3.1 and 3.2, at each pressure level the water-dominated case has a greater temperature difference from dayside to nightside than the 50$\times$ solar case.  Therefore, we expect the water-dominated case to exhibit larger flux variations with orbital phase as compared to the 50$\times$ solar case (Figure \ref{daynight_diff}).  However, this will vary widely as a function of wavelength, as the water spectrum is dominated by fundamental and combination vibrational bands in the near-infrared (IR) and mid-IR.  Within these water bands, the atmosphere is opaque, and hence observations at these wavelengths will probe lower pressures.  Thus, according to Figure \ref{daynight_diff}, we would expect greater day-night temperature variations and larger flux variation with orbital phase.  At wavelengths outside of the water bands (i.e., in spectral windows) observations sense deeper, hotter regions of the atmosphere where day-night temperature variations are smaller; hence, there should be less flux variation with orbital phase.  

\begin{figure}
\begin{centering}
\epsscale{.80}
\includegraphics[trim = 0.0in 0.0in 0.0in 0.0in, clip, width=0.450\textwidth]{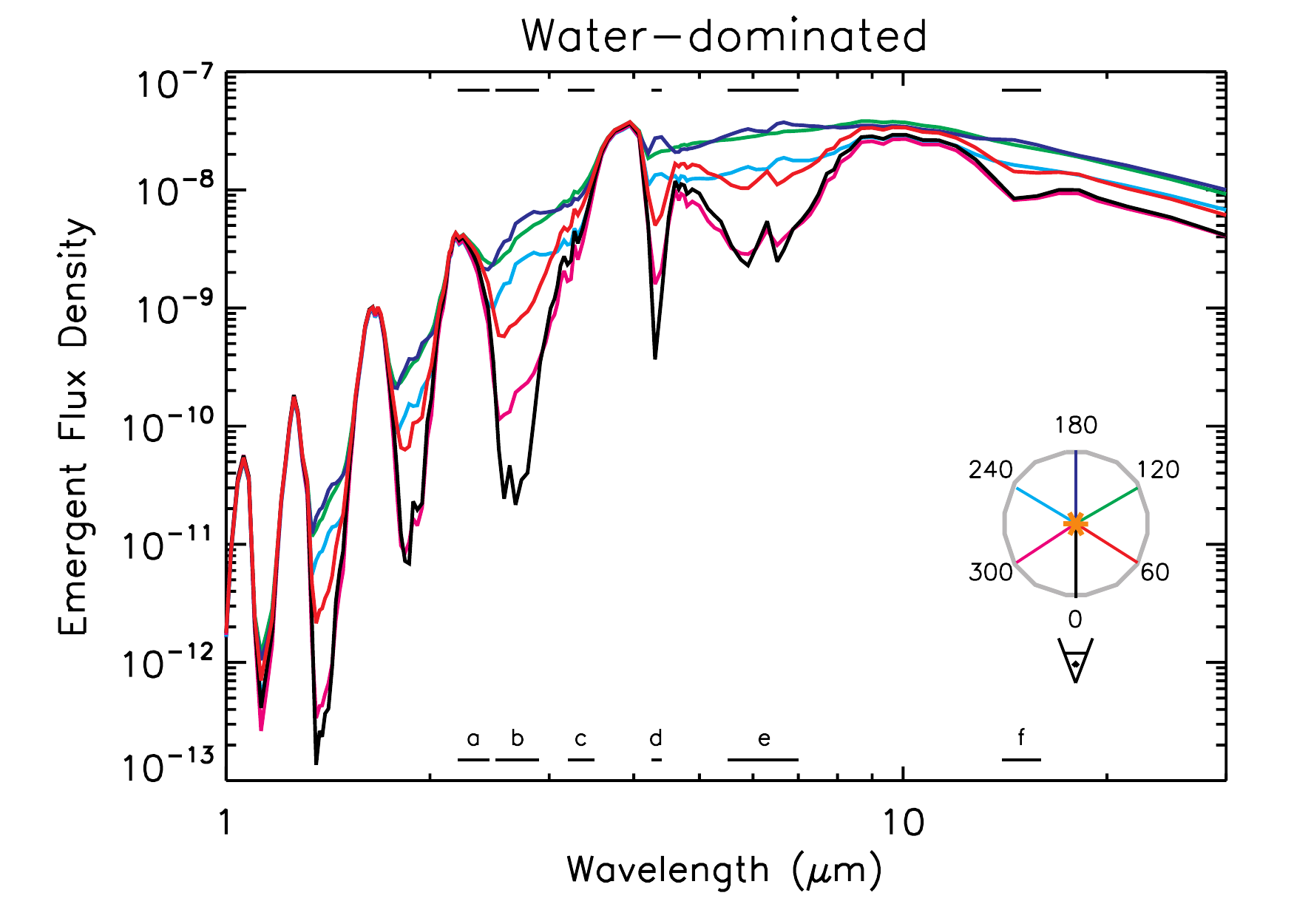}
\includegraphics[trim = 0.0in 0.0in 0.0in 0.0in, clip, width=0.450\textwidth]{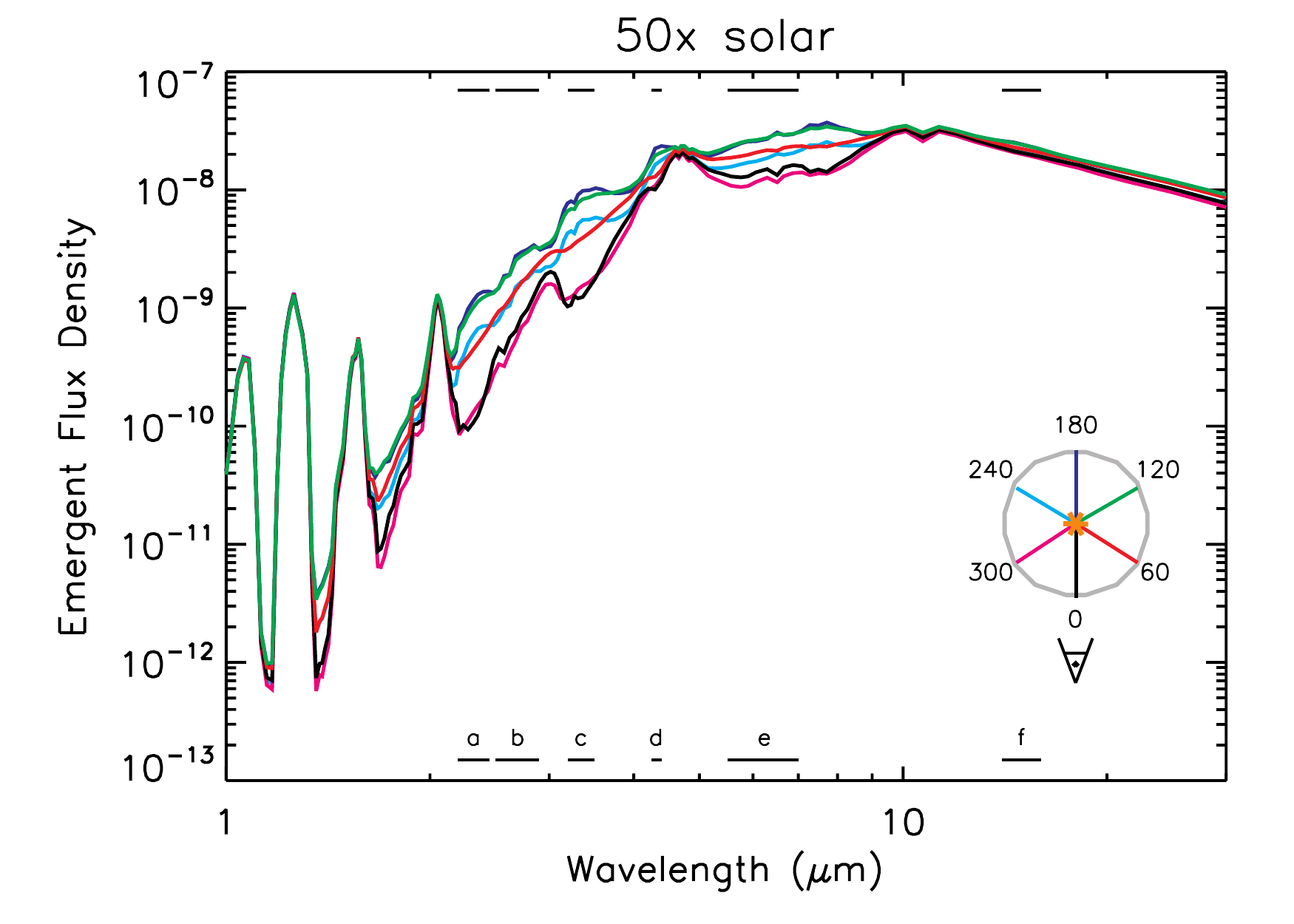}\\
\caption{Emergent flux density (in units of ergs$^{-1}$cm$^{-2}$Hz$^{-1}$) for water-dominated (top panel) and 50$\times$ solar (bottom panel) compositions at six orbital phases: transit, when nightside is visible (black line); 60$^{\circ}$ after transit (red line); 120$^{\circ}$ after transit (green); secondary eclipse, when the dayside is visible (dark blue); 60$^{\circ}$ after secondary eclipse (light blue); and 120$^{\circ}$ after secondary eclipse (magenta).  These phases are illustrated in the inset figure, shown in the bottom right of each panel.  Black horizontal lines indicate the wavelength bands chosen for lightcurves plotted in Figure \ref{lightcurves}, from band $a$ to band $f$. } 
\label{spectra}
\end{centering}
\end{figure}

We see this behavior in theoretical emergent flux spectra for the water-dominated model (Figure \ref{spectra}, top panel).  For both the water-dominated case and the 50$\times$ solar case (bottom panel), the spectra are plotted at six orbital phases, from transit, where the nightside is visible (black spectra), through to 120$^{\circ}$ after secondary eclipse (magenta spectra).  The deep absorption features seen in the water-dominated case are fundamental vibrational bands of water vapor at 2.66, 2.73 and 6.27 microns, as well as combination bands at 1.13, 1.38, 1.88, and 2.68 microns.  Inside the water bands where we are probing low pressure regions, the large day-night temperature variations lead to large flux variations with orbital phase.  Outside of the bands (inside the spectral windows), we probe deeper pressures where there are small temperature differences and hence small phase variations. Comparing the emergent flux spectra of the 50$\times$ solar composition, we see the absorption features are not as deep and less dominated by water features.  The difference between windows and non-windows is also less prominent. However, the 50$\times$ solar case also exhibits variations in emergent flux with phase and wavelength, indicating that for both cases, different atmospheric pressure levels are probed at different wavelengths.  

\begin{figure}
\begin{centering}
\epsscale{.80}
\includegraphics[trim = 0.0in 0.0in 0.0in 0.0in, clip, width=0.5\textwidth]{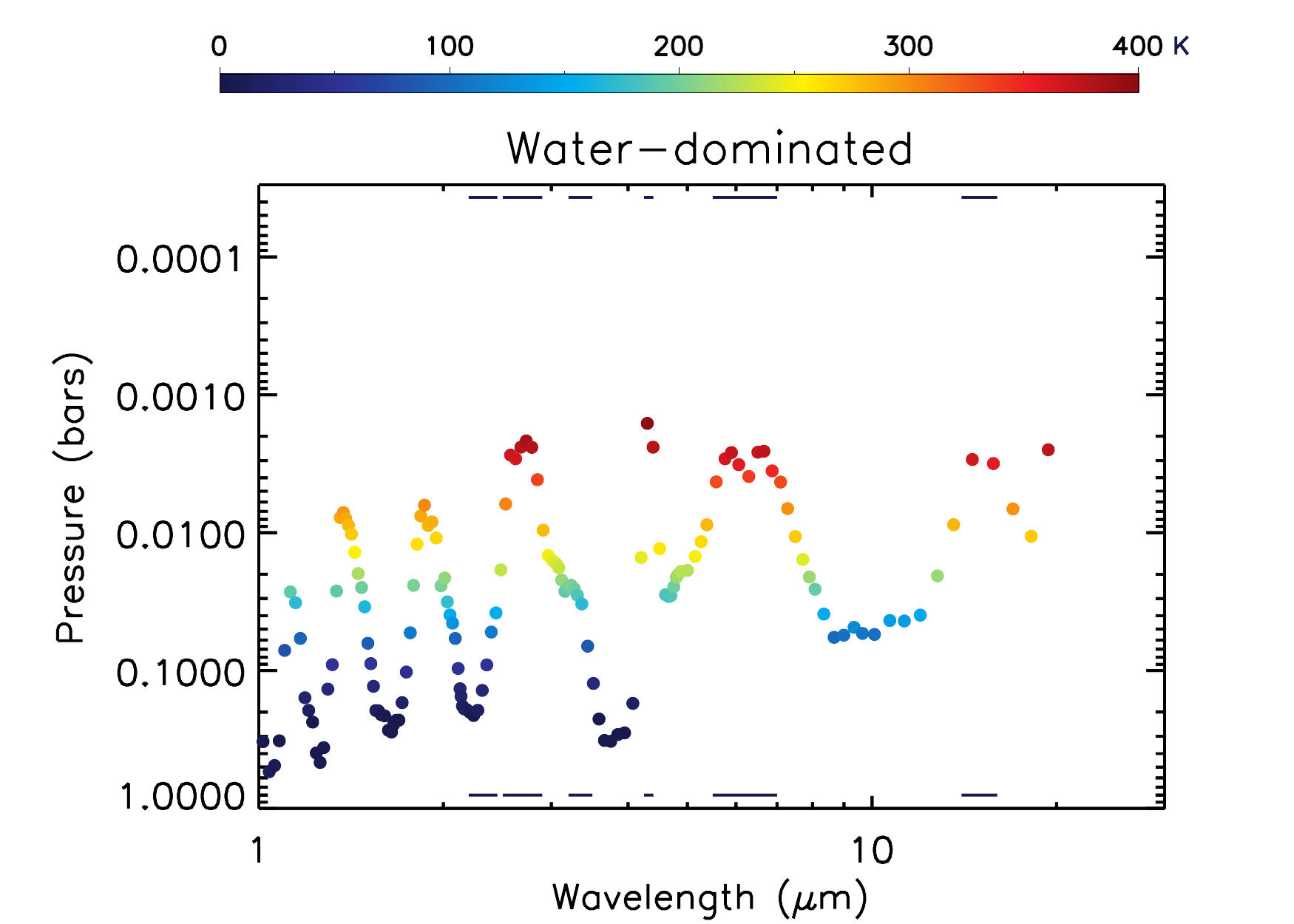}
\includegraphics[trim = 0.0in 0.0in 0.0in 0.0in, clip, width=0.5\textwidth]{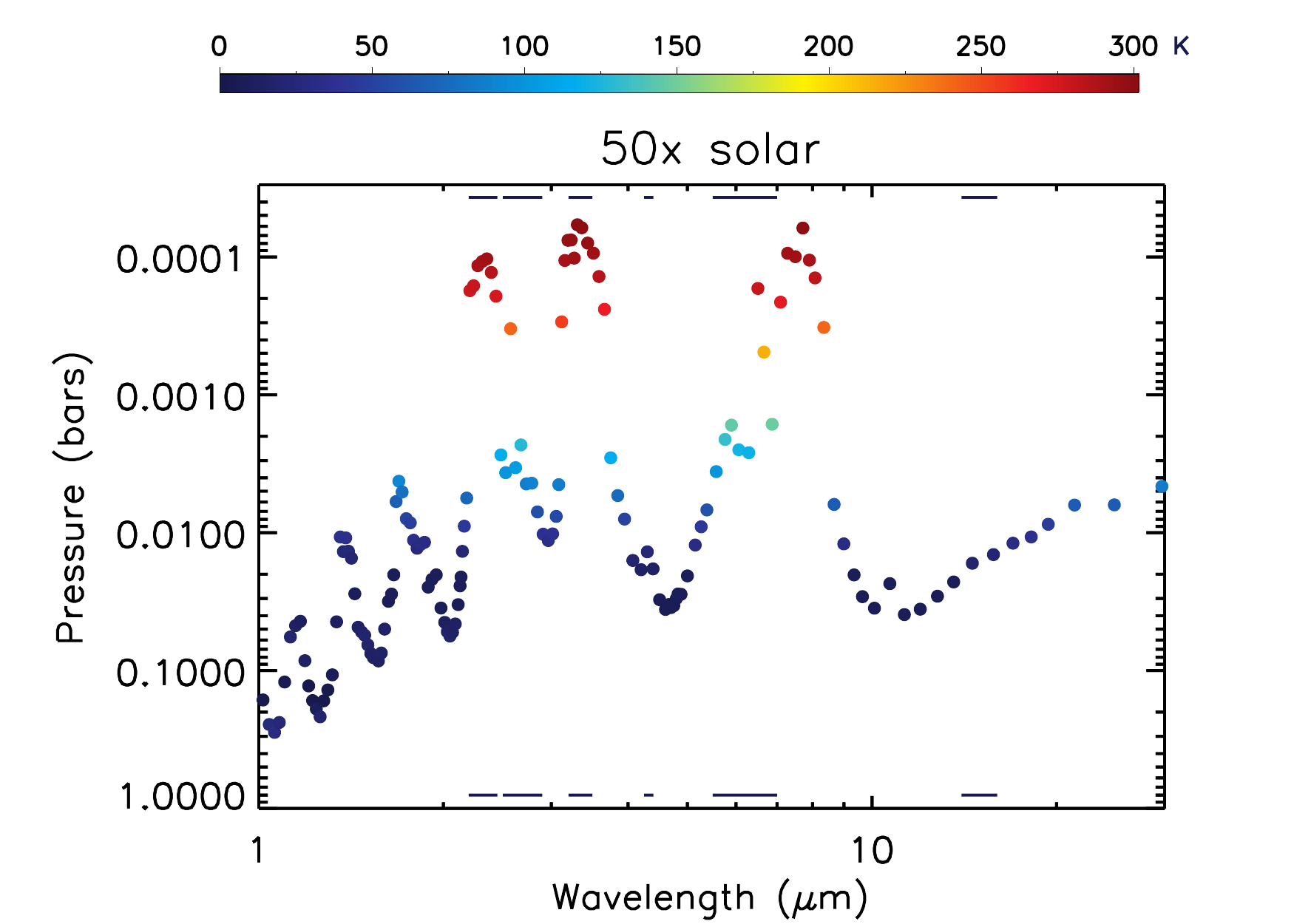}\\
\caption{$\tau=1$ pressure level as a function of wavelength for the water-dominated (top) and 50$\times$ solar (bottom) cases.  The colorscale corresponds to the maximum temperature variation from dayside to nightside, as plotted in Figure \ref{daynight_diff}.  Black horizontal lines indicate the wavelength bands chosen for lightcurves plotted in Figure \ref{lightcurves}, from band $a$ to band $f$. } 
\label{tau1}
\end{centering}
\end{figure}

This can be quantitatively shown by plotting the pressure probed in emergent flux, where the optical depth, $\tau$, is equal to one.  We calculate the $\tau=1$ level by first determining the brightness temperature, $\mathrm{T_{bright}}$, as a function of wavelength, and finding the pressure level at which the globally-averaged temperature is equal to $\mathrm{T_{bright}}$.  The results are plotted in Figure \ref{tau1} with a colorscale corresponding to the maximum temperature variation at each pressure level from Figure \ref{daynight_diff}.  For both atmospheric compositions, the wavelength regions with small (large) phase variations correspond to deeper (shallower) pressures, where day-night temperature variations are smaller (larger).  

\begin{figure}
\begin{centering}
\epsscale{.80}
\includegraphics[trim = 0.0in 0.0in 0.0in 0.0in, clip, width=0.450\textwidth]{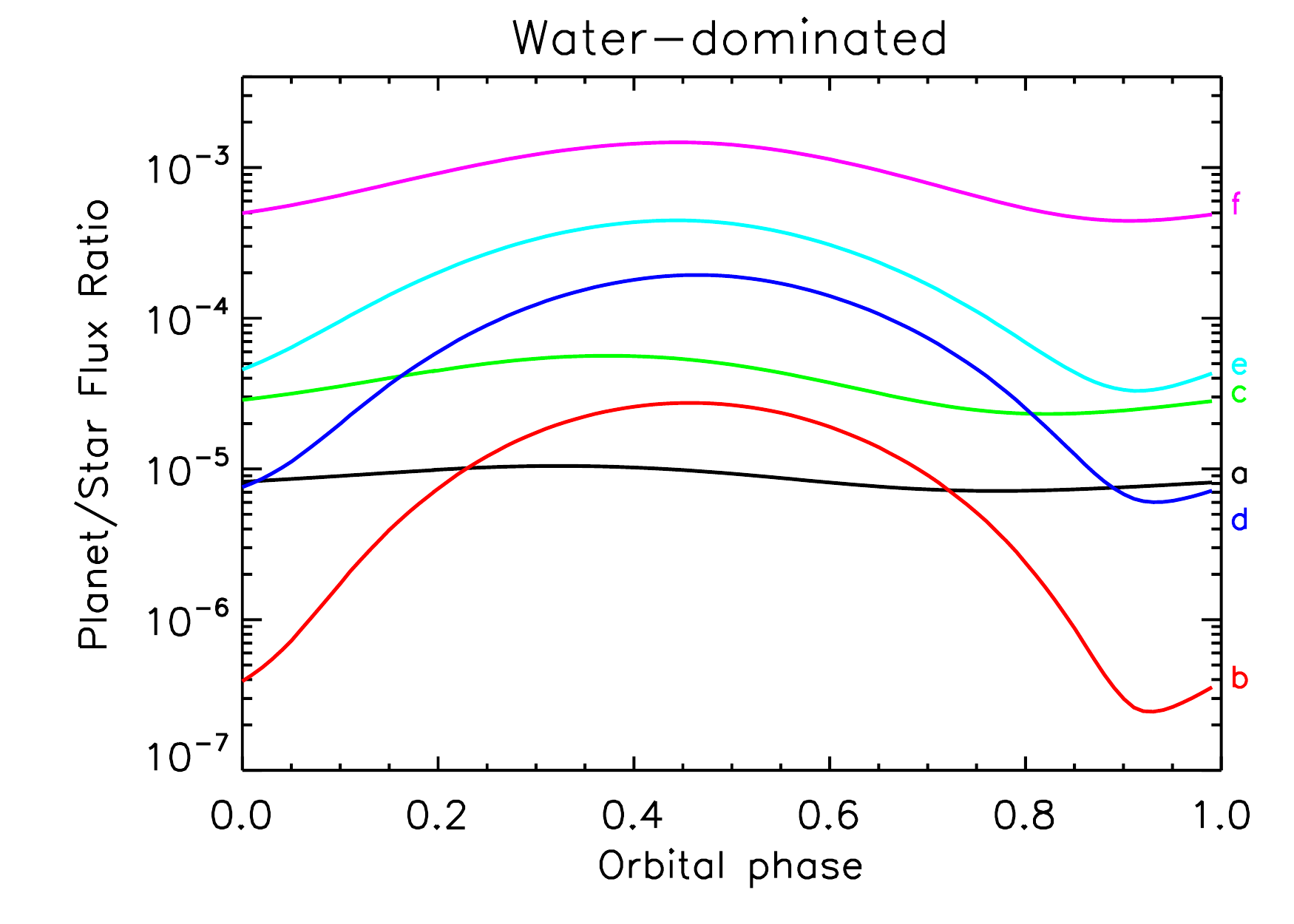}
\includegraphics[trim = 0.0in 0.0in 0.0in 0.0in, clip, width=0.450\textwidth]{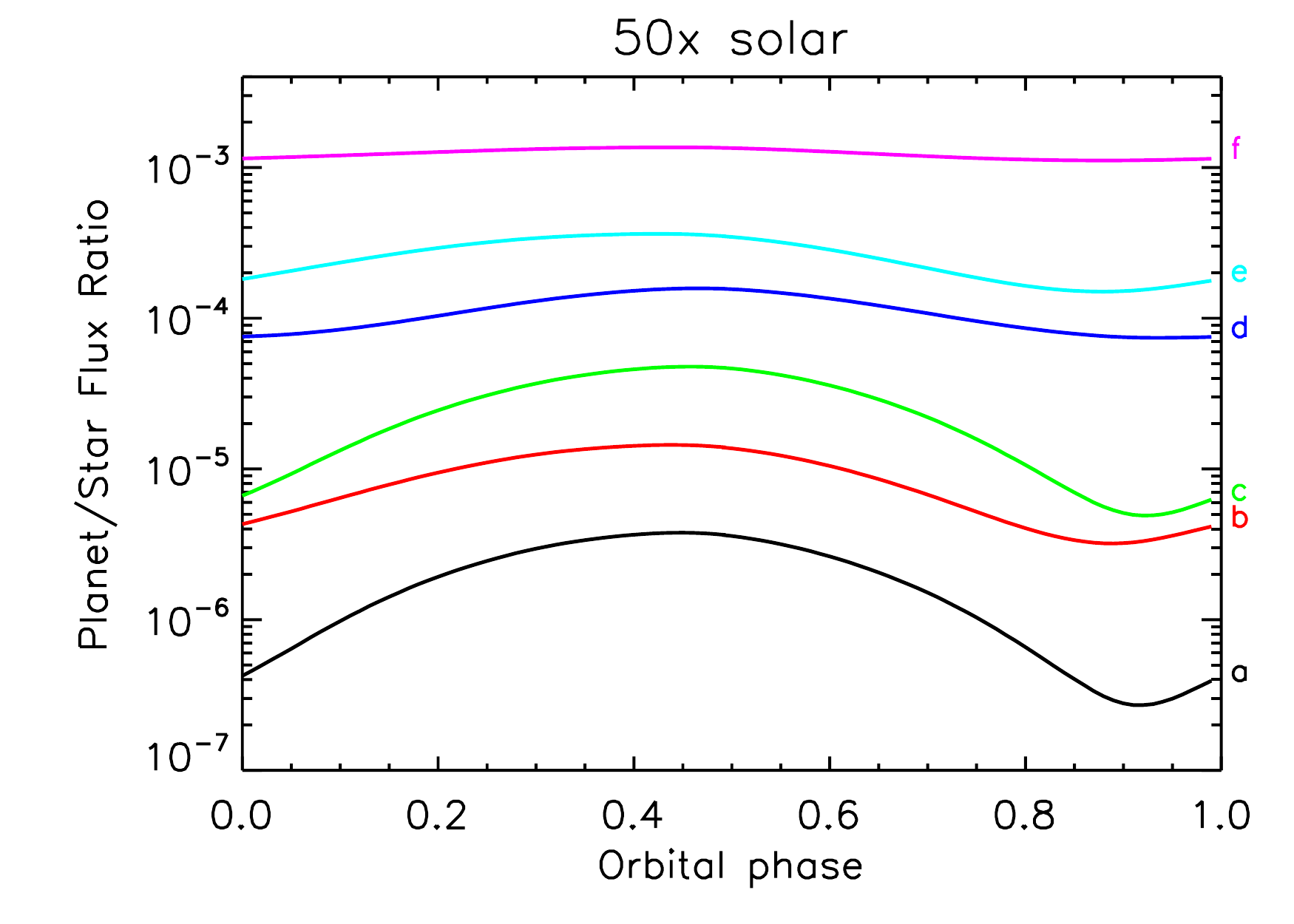}
\caption{Lightcurves plotted as a function of orbital phase for water-dominated (top) and 50$\times$ solar (bottom) atmospheric compositions.  Transit occurs at an orbital phase of 0.0, while secondary eclipse occurs at a orbital phase of 0.5. Each pair of lightcurves correspond to the wavelength bands shown in Figure \ref{spectra}: from bottom to top, bands $a$ (2.2-2.45 $\mu$m, black), $b$ (2.5-2.9 $\mu$m, red), $c$ (3.2-3.5 $\mu$m, green), $d$ (4.25-4.4 $\mu$m, dark blue), $e$ (5.5-7.0 $\mu$m, light blue), and $f$ (14.0-16.0 $\mu$m, magenta).  Note the lightcurves have a bell-shape because they are plotted on a log scale. }
\label{lightcurves}
\end{centering}
\end{figure}

Lightcurves of each composition further illustrate the differences between low- and high-MMW compositions.  We plot the planet/star flux ratio as a function orbital phase for the water- and 50$\times$ solar compositions in Figure \ref{lightcurves}.  In each case, an orbital phase of 0.0 corresponds to transit, while an orbital phase of 0.5 corresponds to secondary eclipse.  Six lightcurves are plotted at the general wavelength bands $a-f$ listed in Table \ref{wl_bands} and denoted in Figures \ref{spectra} and \ref{tau1}.  For the water-dominated case, flux variations are large in all but one band (band $a$, black line). At this band the $\tau=1$ level corresponds to a pressure level of 0.1 bars, where day-night temperature variations are small.  All other wavelength bands probe high in the atmosphere, where day-night temperature variations are large (Figure \ref{tau1}).  For a 50$\times$ solar composition, the flux variations are large for bands $a$, $b$, and $c$ which probe low-pressure regions where day-night temperature variation is high.  Bands $d$, $e$ and $f$ probe deeper pressures, and hence exhibit smaller phase variations.  

Our results demonstrate that one can break the degeneracy in determining the atmospheric composition of GJ 1214b by observing the planet in thermal emission.  Large phase variations within water absorption bands and small variations in its spectral windows would indicate a water-dominated atmosphere.  Other high-MMW species that are highly absorbing, such as carbon dioxide, ammonia, or methane, might in principle exhibit their own characteristic pattern of lightcurve amplitude with wavelength, depending on the wavelengths of their absorption bands and spectral windows.  As shown in Figure \ref{spectra} and \ref{lightcurves}, a hydrogen-dominated atmosphere should exhibit a pattern of lightcurve amplitude with wavelength that differs significantly from that of a high-MMW atmosphere such as one that is water-dominated.  

While we present this method in a generalized sense, one should be able to utilize space-based instruments such as the Near-Infrared Spectrograph (NIRSpec) aboard JWST or the Wide Field Camera 3 (WFC3) on HST, although the latter has less spectral coverage and would require a multitude of orbits to achieve sufficient signal-to-noise.  Instruments on the next generation of ground-based telescopes might also be able to utilize this technique, such as the near-infrared spectrometer (GMTNIRS) on GMT or the Infrared Multi-object Spectrometer (IRMS) on TMT.  However, full-phase lightcurves would be difficult to obtain from the ground in a single observation, and one would have to contend with the water vapor in Earth's atmosphere.  Therefore, reduction of ground-based observations would be much more difficult.  In either case, in order to probe inside and outside water bands effectively as the method requires, spectral observations are necessary.  Observations in photometric passbands (i.e., broadband observations like those on the {\it Spitzer Space Telescope}) might be able to apply this method, but would smear out these spectral features.

These results are particularly favorable because they would generally be independent of the the presence of clouds, minor equilibrium condensates or photochemical haze.  In transit, slant optical depths through the planet's terminator can be dozens of times larger than vertical optical depths \citep{fortney2005a}, which can suppress absorption features.  In emission, however, paths are closer to vertical, suggesting that it is much easier for hazes to flatten the transmission spectrum than the emission spectrum.  Still, if the clouds or hazes are sufficiently thick, they would absorb and scatter the emergent flux which could in turn suppress emission features and flux phase variations.  Given recent \cite{kreidberg2014} results that GJ 1214b likely has clouds or hazes, future work will include exploring how clouds with varying compositions and particle sizes as well as photochemical hazes can affect the phase variations presented here.  

\begin{deluxetable}{lcc}
\tabletypesize{\scriptsize}
\tablecaption{Wavelength bands shown in Figures \ref{spectra}, \ref{tau1} and \ref{lightcurves} in units of microns. \label{wl_bands}}
\tablewidth{0pt}
\tablehead{
\colhead{Wavelength band} & \colhead{Left bound ($\mu$m)} & \colhead{Right bound ($\mu$m)}
}
\startdata
$a$ & 2.2 & 2.45\\
$b$ & 2.5 & 2.9\\
$c$ & 3.2 & 3.5\\
$d$ & 4.25 & 4.4 \\
$e$ & 5.5 & 7.0 \\
$f$ & 14.0 & 16.0\\
\enddata
\end{deluxetable}

\section{Conclusions}
We present three-dimensional atmospheric circulation models of the super-Earth GJ 1214b, exploring changes in circulation as a function of metallicity and composition.  For hydrogen-dominated atmospheres, atmospheric opacities are enhanced with increasing metallicity, leading to shallower atmospheric heating.  This yields strong dayside-nightside heating/forcing that increases with metallicity, which in turn produces the highest day-night temperature variations and hence the strongest equatorial superrotation in the 50$\times$ solar model.

The water-dominated composition also exhibits superrotation at the equator and eastward jets at high latitudes, but the circulation of the $\mathrm{CO_2}$-dominated model is dominated mainly by high-latitude jets. All three high-MMW models have higher horizontal temperature variations at a given (low) pressure than the low-MMW models.  These differences in temperature structure and circulation can be attributed to differences in opacity structure and scale height. 

The theoretical dayside lightcurves and spectra presented here lead to a major prediction for how to break the current observational degeneracy in the composition of GJ 1214b's atmosphere. In particular, the water bands dominate the spectra of the 99\% $\mathrm{H_2O}$, 1\% $\mathrm{CO_2}$ case.  Within water absorption bands, large day-night temperature variations lead to large flux variations with phase.  Outside of the water bands (within atmospheric windows), these phase variations are small.  In comparison, a 50$\times$ solar atmosphere generally yields small phase variations at those wavelength bands and large phase variations at other characteristic bandpasses.  Therefore, observing in emission would break the degeneracy to determining the atmospheric composition of GJ 1214b.  One could potentially constrain the existence of water or other highly-absorbing species by selecting wavelength bands inside and outside of their atmospheric windows, and comparing the extent of phase variations with that of a low-MMW atmosphere.  This diagnostic is much less sensitive than transit spectra to clouds, condensates and hazes.  However, sufficiently thick clouds and hazes would absorb and scatter emergent flux, therefore diminishing emission features and flux variations with orbital phase.

\acknowledgments
This work was supported by Origins grant NNX12AI79G to APS.  T.K. also acknowledges support from the Harriet P. Jenkins Pre-Doctoral Fellowship Program (JPFP).  Resources supporting this work were provided by the NASA High-End Computing (HEC) Program through the NASA Advanced Supercomputing (NAS) Division at Ames Research Center.  We thank Roxana Lupu for comments and for providing opacity tables for $\mathrm{CO_2}$-$\mathrm{CO_2}$ pressure-induced absorption.  We also thank Daniel Apai, Nikole Lewis, Robert Zellem for insightful discussions.  Lastly, we thank the anonymous referee for their helpful comments and suggestions.

\clearpage

\end{document}